\documentclass[lettersize,journal]{IEEEtran}
\usepackage{amsfonts,bm,booktabs,cite,graphicx,hyperref,multirow,textcomp,url}
\usepackage{amsmath}
\usepackage{algorithmic}
\usepackage{algorithm}
\usepackage{array}
\usepackage[caption=false,font=normalsize,labelfont=sf,textfont=sf]{subfig}
\usepackage{textcomp}
\usepackage{stfloats}
\usepackage{url}
\usepackage{verbatim}
\usepackage{graphicx}
\usepackage{cite}
\usepackage{bbding}
\usepackage[flushleft]{threeparttable}
\begin{document}

\title{URVC: A Unified Real-Time Neural Video Coding Model with Temporal, Spatial, and Perceptual Adaptivity}

\author{Xihua Sheng,~\IEEEmembership{Member,~IEEE}, 
Chang Wen Chen,~\IEEEmembership{Life Fellow,~IEEE}\\
        % <-this % stops a space
\thanks{Date of current version \today.}
\thanks{X. Sheng and C. Chen are with the Department of Computing, The Hong Kong Polytechnic University, Hong Kong SAR, China (e-mail: xihua.sheng@polyu.edu.hk, changwen.chen@polyu.edu.hk).}

\thanks{Corresponding Author: Chang Wen Chen.}% <-this % stops a space
}

% The paper headers
\markboth{Journal of \LaTeX\ Class Files,~Vol.~14, No.~8, August~2021}%
{Shell \MakeLowercase{\textit{et al.}}: A Sample Article Using IEEEtran.cls for IEEE Journals}

%\IEEEpubid{0000--0000/00\$00.00~\copyright~2021 IEEE}

% Remember, if you use this you must call \IEEEpubidadjcol in the second
% column for its text to clear the IEEEpubid mark.

\maketitle

\begin{abstract}
Neural video coding has advanced rapidly, achieving competitive compression performance while also enabling real-time coding speed. 
Yet, existing codecs exhibit severe rigidity when deployed in dynamic environments, failing to adapt to different video content, user requirements, and quality preferences.
First, to meet the real‑time constraint, they discard explicit motion estimation and motion compression, thereby losing the ability to adapt temporal prediction to motion complexity and bitrate constraints.
Second, their spatial bit allocation strategy is coarse and, once trained, is fixed. It cannot adapt to dynamic user requirements at test time, preventing users from freely controlling the spatial distribution of bits.
Third, they cannot adapt their quality preference to varying application requirements without deploying separate models.
We address all three limitations within a single real-time neural video codec---URVC, transforming a rigid system into a unified framework with temporal, spatial, and perceptual adaptivity.
First, we propose a rate-aware adaptive temporal prediction method that generates diverse prediction candidates through a multi-candidate architecture and couples candidate selection directly to rate‑distortion optimization.
Second, we propose a decomposition-based spatial rate control method that achieves finer‑grained spatial bit allocation through feature decomposition and separate quantization, and allows users to perform direct spatial rate control at test time without retraining.
Third, we propose a perceptual switching method that only requires learning a secondary module bank alongside a frame generator, enabling a codec to switch between signal fidelity and perceptual quality modes.
Extensive experiments validate that URVC surpasses the state-of-the-art real-time codec DCVC-RT in terms of YUV420 PSNR under intra period $-1$, delivers substantially better perceptual quality, and supports zero-shot region‑of‑interest (ROI) coding.
These results confirm that the proposed methods substantially enhance the temporal, spatial, and perceptual adaptivity of neural video coding, paving the way for practical deployment in real-world applications.

\end{abstract}


\section{Introduction}
Video has become the dominant form of Internet traffic, continuously driving the demand for more efficient video compression. Traditional video codecs, represented by H.264/AVC, H.265/HEVC, and H.266/VVC, achieve remarkable compression performance through a highly adaptive hybrid coding framework. Instead of relying on a fixed coding strategy, the encoder dynamically selects among multiple coding modes according to the content characteristics and rate–distortion optimization. For example, it adaptively chooses motion prediction modes to accommodate diverse motion patterns, adjusts quantization parameters to allocate bits across spatial regions according to both content characteristics and user requirements, and switches between fidelity-oriented and perception-oriented coding configurations under different application scenarios. Such content- and user-adaptive decision making forms the foundation of modern video compression, enabling traditional codecs to flexibly balance coding efficiency, visual quality, and deployment requirements across diverse scenarios.

In recent years, the pursuit of higher compression performance has driven the emergence of neural video coding (NVC)~\cite{lu2020end,sheng2025prediction,hu2022fvc,agustsson2020scale,cheng2019learning,rippel2019learned,liu2022end,yilmaz2021end,lin2022dmvc,guo2023learning, liu2020conditional,wei2025rdvc,jia2025towards,yuan2025mining,ho2022canf,chen2024maskcrt,sheng2022temporal,sheng2024vnvc,sheng2025bi,li2023neural,sheng2024spatial,li2024neural}. NVC replaces key components of the hybrid video coding framework---motion estimation, motion compression, motion compensation, residual coding, and entropy modeling---with neural networks, and trains the entire model end-to-end under a rate-distortion loss. Through architectural innovations such as scale-space flow~\cite{agustsson2020scale}, feature-space motion estimation and compensation~\cite{hu2022fvc}, temporal context mining~\cite{sheng2022temporal}, and quadtree-based entropy models~\cite{li2023neural}, neural video codecs have surpassed the compression performance of traditional standards. However, this performance gain has come at the cost of extreme computational complexity---many NVC models exceed one million multiply-accumulate operations (MACs) per pixel, making them impractical for real-world deployment. To further bridge the gap between compression efficiency and practical deployment, the recently proposed real-time neural video codec DCVC-RT~\cite{jia2025towards} removes explicit motion estimation, motion compression, and motion compensation modules, replacing them with lightweight implicit temporal modeling. This simplified architecture significantly reduces computational complexity while preserving competitive compression performance, making real-time neural video coding practically feasible.\par

Despite the remarkable progress in real-time neural video coding, pursuing efficiency comes at the cost of adaptivity. Unlike traditional codecs, whose coding decisions are dynamically optimized for every frame according to content characteristics and rate--distortion optimization, a real-time neural codec performs almost all coding decisions implicitly through a fixed neural network learned offline. Once training is completed, its coding strategies are fixed and cannot adapt to the characteristics of the current video and user requirements. 
As a consequence, several adaptive capabilities that have long been fundamental to video compression are inherently lost.

First, temporal prediction can no longer adapt to diverse motion patterns and bitrate constraints. Traditional codecs provide rich prediction diversity by evaluating multiple motion hypotheses and selecting the one that minimizes the rate--distortion cost. Earlier neural video codecs retain a similar capability through explicit motion estimation, motion compression, and motion compensation, where the prediction quality can be adjusted by allocating different numbers of bits to motion information. In contrast, recent real-time neural codecs replace explicit motion modeling with an implicit prediction network. Such a fixed mapping generates temporal context in exactly the same manner regardless of motion complexity or bitrate, making it difficult to accommodate diverse motion patterns or dynamically balance prediction accuracy under different rate constraints. 
\par

Second, spatial bit allocation cannot adapt to  user requirements. Although existing neural video codecs learn spatially adaptive bit allocation through end-to-end optimization, the learned allocation strategy becomes fixed once training is completed. Unlike traditional codecs, where users can freely adjust quantization parameters for different coding units to satisfy application-specific requirements such as region-of-interest (ROI) coding~\cite{chen2003roi}, neural codecs provide almost no flexibility for user-controlled spatial bit allocation at deployment. Existing ROI-oriented neural codecs~\cite{fathima2023neural,wu2024roi,liu2024roi} attempt to recover this capability by introducing ROI masks or ROI-aware supervision during training. However, these methods generally require ROI annotations or ROI-aware supervision, complicate the training procedure, and may compromise reconstruction quality outside the selected regions. Consequently, current neural codecs still lack a general mechanism for user-driven spatial rate control without retraining.\par

Third, the codec cannot adapt its quality preference to varying application requirements. Existing neural video codecs are typically optimized for a single quality preference throughout training, causing each model to specialize in either signal fidelity or perceptual quality. Fidelity-oriented codecs achieve excellent objective reconstruction quality but tend to produce overly smooth textures at low bitrates, whereas perceptual codecs generate visually realistic details at the expense of signal fidelity, with fidelity remaining substantially low even at high bitrates. As application requirements vary continuously in practice, different quality preferences often coexist within the same deployment environment. However, satisfying these different objectives usually requires maintaining multiple independently trained codecs, which substantially increases storage, maintenance, and deployment costs. A practical neural video codec should therefore support flexible quality adaptation within a single model.
\par

Motivated by the above observations, we propose URVC, a unified real-time neural video codec that achieves temporal, spatial, and perceptual adaptivity within a single lightweight framework. First, we propose a rate-aware adaptive temporal prediction method that generates diverse prediction candidates through a multi-candidate architecture and couples candidate selection directly to rate‑distortion optimization. This enables temporal prediction to adapt jointly to motion complexity and bitrate constraints without explicit motion estimation. Second, we propose a decomposition-based spatial rate control method that achieves finer-grained spatial bit allocation through feature decomposition and separate quantization, and allows users to perform direct spatial rate control at test time according to their requirements. This enables zero-shot ROI coding without retraining or additional supervision. Third, we propose a perceptual switching strategy that allows a neural video codec to switch between signal-fidelity and perceptual-quality modes according to different quality preferences, eliminating the need for multiple separately trained models.

The main contributions are summarized as follows:
\begin{itemize}
\item We revisit neural video coding from the perspective of adaptive coding and identify a fundamental limitation of existing neural codecs: although they achieve remarkable compression performance, their coding strategies remain largely fixed at deployment, limiting adaptivity in temporal prediction, spatial rate allocation, and quality preference.

\item We propose URVC, a unified real-time neural video codec that simultaneously achieves temporal, spatial, and perceptual adaptivity, transforming a fixed-strategy codec into a deployment-adaptive coding framework.

\item We develop a rate-aware adaptive temporal prediction method, a decomposition-based spatial rate control method, and a perceptual switching method, which collectively enable adaptive temporal prediction, user-controllable spatial rate allocation, and allow a neural codec to serve both signal fidelity and perceptual quality.

\item Extensive experimental results show that our codec surpasses the state-of-the-art real-time codec DCVC‑RT in terms of YUV420 PSNR under intra period $-1$, delivers substantially better perceptual quality, and achieves superior compression performance on video motion control, an emerging ROI coding task, without compromising the video generation quality.
\end{itemize}

The remainder of this paper is organized as follows. Section~\ref{related_work} reviews related work. Section~\ref{sec:method} describes our proposed methods. Section~\ref{sec:experiment} presents experimental results. Section~\ref{sec:conclusion} concludes the paper.

\section{Related Work}\label{related_work}
\subsection{Temporal Prediction for Neural Video Coding}
Temporal prediction, which removes inter‑frame redundancy by leveraging the information of previously decoded frames, is a cornerstone of video compression. In traditional codecs, this is achieved through explicit motion estimation and compensation that can generate multiple prediction candidates and select the optimal one under rate‑distortion constraints. Early neural video codecs inherit this philosophy but replace hand‑crafted modules with learnable networks. The pioneering DVC~\cite{lu2020end} introduces an end‑to‑end framework where an optical flow network estimates pixel‑wise motion vectors, which are then compressed via an auto‑encoder and used for bilinear warping‑based motion compensation. This motion‑estimation‑compression‑compensation pipeline has become the dominant paradigm for temporal prediction in NVCs.

Since DVC, many works have advanced this paradigm by improving motion representation and alignment. Scale‑space flow~\cite{agustsson2020scale} augments the conventional two‑channel optical flow with a per‑pixel scale field, allowing the warping operation to fall back to a blurred prediction when motion is unreliable. FVC~\cite{hu2022fvc} moves all operations---motion estimation, compression, and compensation---into a feature space and uses deformable convolutions for more flexible alignment.  DCVC‑TCM~\cite{sheng2022temporal} propagates decoded features instead of decoded frames and learns multi‑scale temporal contexts. Despite their success in rate‑distortion performance, these codecs incur enormous complexity---often exceeding thousands of MACs per pixel---due to the many layers required for motion estimation, compression, and compensation. The high computational cost makes real‑time deployment impractical.

To break this complexity barrier, DCVC‑RT~\cite{jia2025towards} discards the entire explicit motion branch, including motion estimation, motion compression, and warping-based motion compensation. Instead, temporal contexts are generated by a lightweight implicit feature extractor composed solely of depth‑wise convolutions, which operates on a cached reference feature. This design reduces the number of modules drastically and achieves real‑time coding while maintaining competitive rate‑distortion performance. However, the implicit temporal modeling of DCVC‑RT relies on a fixed prediction rule that lacks the adaptivity in motion complexity and the bitrate constraints. 

\subsection{Spatial Rate Control for Neural Video Coding}
In traditional block‑based video codecs, spatial rate control can be achieved by varying the quantization parameter (QP) across coding units, enabling fine‑grained bit allocation that respects region‑of‑interest priorities or content complexity. In existing neural video codecs, such as DCVC‑DC~\cite{li2023neural}, DCVC‑FM~\cite{li2024neural}, and DCVC‑RT~\cite{jia2025towards}, their entropy models learn a spatially varying quantization step, which modulates the latent representation before rounding. This enables some degree of content‑adaptive bit allocation---the quantization step can vary across spatial locations based on local content complexity. However, the quantization step operates at the resolution of the latent representation, which is far below the granularity needed for precise spatial control. Moreover, once trained, this allocation strategy is fixed. At test time, users cannot freely adjust the quantization strength across spatial regions, making it impossible to control the bit allocation according to user requirements.

Some works have explored ROI‑based video compression in this context. Fathima et al.~\cite{fathima2023neural} introduced two ROI‑based neural codecs based on scale‑space flow, where an ROI mask is concatenated to the input and the distortion loss is weighted per region. Wu et al.~\cite{wu2024roi} applied a ROI loss to DVC, using a detection network to generate masks and a resolution‑adaptive inference strategy. Liu et al.~\cite{liu2024roi} proposed ROI‑aware dynamic quantization for NVC, where a bit‑allocator assigns different bit‑widths to ROI and non‑ROI based on motion and texture complexity.
However, these methods require the ROI mask to be provided during training. Most of them achieve high ROI fidelity at the cost of severely degraded non‑ROI reconstruction. When ROI coding is not needed, the performance of these codecs drops significantly below that of a model trained without ROI loss. None of existing codecs support zero‑shot ROI coding at test time. 

\subsection{Perceptual Optimization for Neural Video Coding}
Existing perceptual neural video codecs can be broadly classified into four categories: generative adversarial network (GAN)-based, vector quantization (VQ)-based, diffusion-based, and visual auto‑regressive (VAR)-based.

GAN‑based codecs pioneered perceptual neural video coding. PLVC~\cite{yang2022perceptual} conditions a recurrent discriminator on motion and latent representations to improve spatio‑temporal realism. HVFVC~\cite{li2023high} introduces confidence‑based feature reconstruction and a periodic compensation loss to mitigate checkerboard artifacts. VQ‑based codecs leverage vector‑quantized auto‑encoders for a semantically rich latent space. GLC‑Video~\cite{qi2025generative} performs transform coding in the latent space of a generative VQ‑VAE, achieving ultra‑low bitrate compression with high perceptual fidelity. Diffusion‑based codecs exploit pre‑trained diffusion models as generative priors. DiffVC~\cite{ma2025diffusion} integrates a multi‑step diffusion process into conditional coding. DiffVC‑RT~\cite{ma2026diffvc} reduces inference to a single diffusion step and prunes the backbone for real‑time decoding. GNVC‑VD~\cite{mao2026generative} uses a video diffusion transformer for sequence‑level latent refinement, significantly improving temporal coherence. VAR‑based codecs adopt a next‑scale prediction paradigm. ProGVC~\cite{li2026progvc} unifies progressive scalability and detail synthesis within a single autoregressive pipeline, enabling bitrate‑adaptive perceptual compression.

Despite their impressive perceptual gains, these methods share two fundamental limitations. First, they heavily sacrifice signal fidelity even at high bitrates. Second, their computational complexity remains prohibitively high for practical deployment. For example, the latest diffusion‑based DiffVC‑RT consumes 2462K MACs per pixel. This complexity, combined with the need for two separate models to cover both fidelity and perceptual scenarios, severely limits their practicality in resource‑constrained environments.

%-------------------------------------------------------------------------
\begin{figure*}[t]
  \centering
   \includegraphics[width=\linewidth]{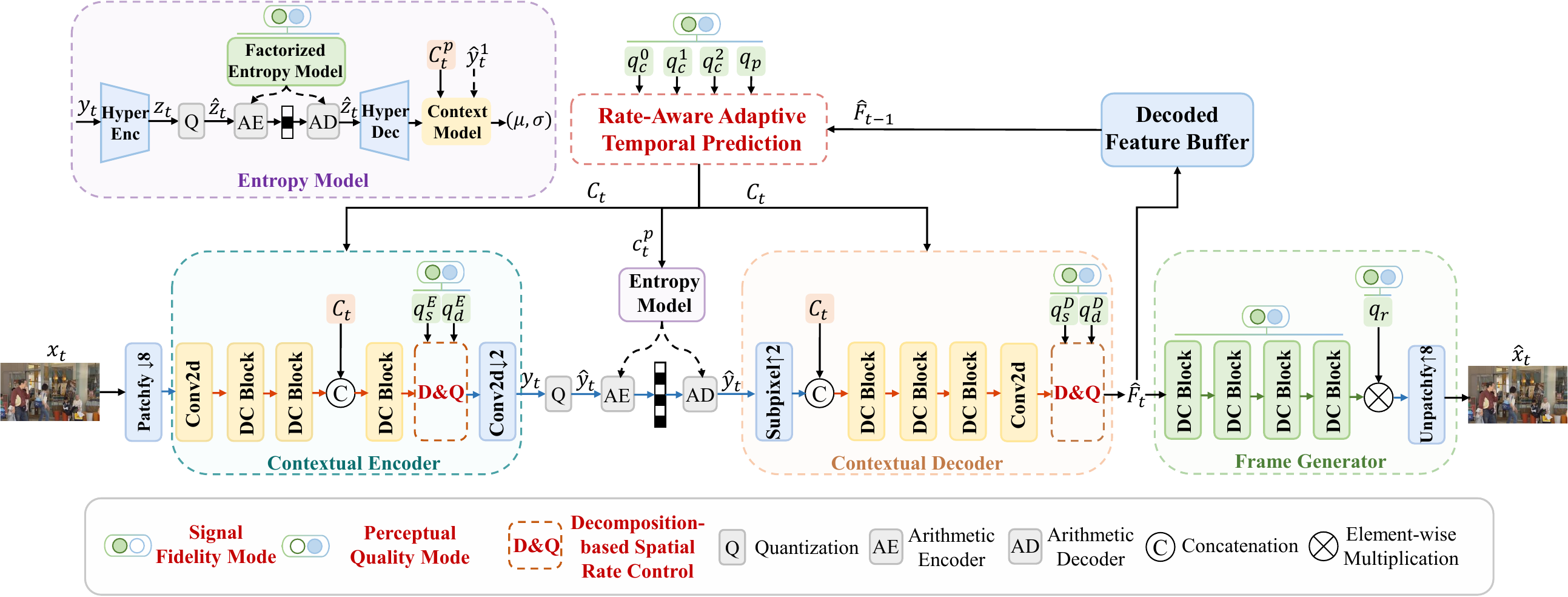}
      \caption{Architecture of our proposed real-time neural video codec with rate-aware adaptive temporal prediction method, decomposition-based spatial rate control method, and perceptual switching method.}
   \label{fig:framework}
\end{figure*}
%-------------------------------------------------------------------------
\section{Method}\label{sec:method}
\subsection{Overview}
We first present an overview of our real‑time neural video codec---URVC, which integrates our proposed adaptive temporal prediction method, spatial rate control method, and perceptual switching method, as illustrated in Fig.~\ref{fig:framework}.

\subsubsection{Rate-Aware Adaptive Temporal Prediction}
To overcome the deterministic nature of implicit temporal prediction, we design a rate-aware adaptive temporal prediction module.
It takes the reference feature $\hat{F}_{t-1}$ and produces a temporal context $C_t$ and an auxiliary temporal prior $C^p_t$.
The module structurally introduces prediction diversity through a multi-branch architecture that generates multiple prediction candidates.
The final context $C_t$ is formed by fusing these candidates, with the learnable quantization vectors serving as the fusion weights.
This couples candidate selection to both the rate constraint and motion complexity, restoring the temporal adaptivity lost in deterministic implicit prediction.
Details are provided in Section~\ref{sec:adaptive_temporal_prediction}.

\subsubsection{Decomposition-based Contextual Encoder and Decoder}
The contextual encoder transforms the input frame $x_t$ into a compact latent representation $y_t$. A pixel‑unshuffle layer first reduces the spatial resolution by a factor of $8$, and the resulting feature is concatenated with the temporal context $C_t$ and processed by depth‑wise convolutional layers. Before the final down‑sampling, the intermediate feature undergoes our proposed spatial decomposition quantization, which enables finer‑grained bit allocation according to content complexity. The contextual decoder mirrors the encoder, up‑sampling the quantized latent $\hat{y}_t$ and applying the same spatial decomposition quantization to its intermediate feature. At test time, the down‑up sampling-based decomposition in the encoder can be replaced by an arbitrary spatial mask, giving users direct spatial rate control. The decoded feature $\hat{F}_t$ is stored in the decoded buffer as the reference for the next frame and forwarded to the frame generator. Details are given in Section~\ref{sec:spatial_rate_control}.

\subsubsection{Frame Generator with Perceptual Switching}
The frame generator refines the decoded feature $\hat{F}_t$ by a series of depth‑wise convolutional layers and performs a pixel‑shuffle operation with an up‑sampling factor of $8$ to produce the reconstructed frame $\hat{x}_t$.
Two separate instances of the frame generator are maintained: $\theta_r^{\text{Fid}}$ for the fidelity mode and $\theta_r^{\text{Per}}$ for the perceptual mode. At inference time, users can switch between the two modes by selecting the corresponding instance.
Further details are provided in Section~\ref{sec:perceptual_switching}.

\subsubsection{Entropy Model}
We use a hyperprior entropy model to provide hyper prior for $y_t$. In parallel, the temporal prior $C^p_t$ is fused with the hyper prior to predict the Gaussian distribution parameters for $y_t$. The quantized side information $\hat{z}_t$ is compressed using a separate factorized entropy model~\cite{jia2025towards}, which is part of the module bank and indexed by the current QP to match the distribution of $\hat{z}_t$ at different bitrates.

\subsubsection{Module Bank with Perceptual Switching}
The learnable quantization vectors that modulate the encoder, decoder, temporal prediction module, and frame generator, together with the factorized entropy model for compressing the side information $\hat{z}_t$, form a module bank in which all components are indexed by QP. To enable perceptual switching, we maintain two distinct module banks: $\mathcal{B}^{\text{Fid}}$ optimized for signal fidelity, and $\mathcal{B}^{\text{Per}}$ trained for perceptual quality. At inference time, users can switch between a fidelity mode and a perceptual mode by simply loading the corresponding module bank. More details can be found in Section~\ref{sec:perceptual_switching}.

\subsection{Rate-Aware Adaptive Temporal Prediction}\label{sec:adaptive_temporal_prediction}
In existing real‑time neural video codecs, temporal context prediction is formulated as a deterministic mapping:
\begin{equation}
C_t = f_\theta(\hat{F}_{t-1}),
\end{equation}
where \(f_\theta\) is a neural network with fixed parameters \(\theta\) and \(\hat{F}_{t-1}\) denotes the reference feature from the previously decoded frame.
Let \(F_t\) be the feature representation of the current frame \(x_t\) (e.g., obtained after the pixel‑unshuffle and initial convolutional layers of the encoder).
From a probabilistic perspective, this implicit temporal modeling assumes that the conditional distribution of \(F_t\) given \(\hat{F}_{t-1}\) can be approximated by a unimodal Gaussian with constant variance:
\begin{equation}
p(F_t \mid \hat{F}_{t-1}) \sim \mathcal{N}\bigl(f_\theta(\hat{F}_{t-1}),\; \sigma^2).
\end{equation}

Real‑world videos, however, exhibit highly non‑stationary temporal statistics due to fast translation, non‑rigid deformation, and occlusion.
When the actual motion deviates from the training distribution, a statistical mismatch occurs, making the empirical prediction variance far exceed the model's assumption:
\begin{equation}
\mathbb{E}\bigl[\|F_t - f_\theta(\hat{F}_{t-1})\|^2\bigr] \gg \sigma^2.
\end{equation}
Because the fixed mapping can adapt neither to motion complexity nor to bitrate constraints, it may provide adequate prediction under simple motion, but can lead to substantially increased residuals under complex motion or at low bitrates, rendering temporal prediction potentially fragile.

To overcome the statistical mismatch caused by the deterministic unimodal assumption, we draw on the bias‑diversity trade‑off from ensemble learning.
According to the Ambiguity Decomposition theorem~\cite{krogh1994neural}, the mean squared error (MSE) of an ensemble \(\bar{f} = \sum_{k=0}^{K-1} w_k \phi_k\) (with \(\sum_k w_k = 1\)) can be rigorously decomposed as
\begin{equation}
\text{MSE}(\bar{f}) = \sum_{k=0}^{K-1} w_k \text{MSE}(\phi_k) - \sum_{k=0}^{K-1} w_k \|\phi_k - \bar{f}\|^2,
\label{eq:4}
\end{equation}
where the first term represents the average individual error and the second term quantifies the ensemble \emph{diversity} (or ambiguity).
It indicates that to reduce the total prediction error, it is beneficial to construct a set of heterogeneous base estimators that possess high diversity, thereby increasing the second term.
Homogeneous branches would substantially reduce the diversity term, limiting the ensemble's ability to capture multimodal dependencies.
%-------------------------------------------------------------------------
\begin{figure}[t]
  \centering
   \includegraphics[width=0.9\linewidth]{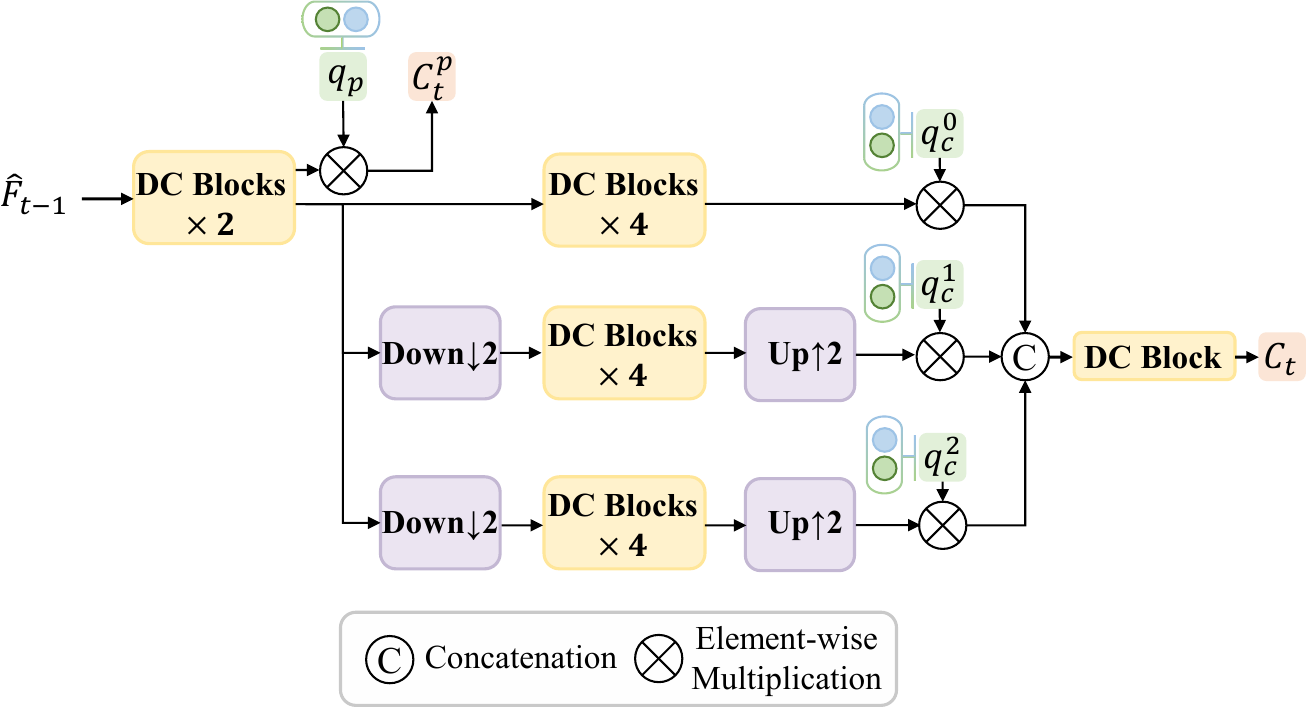}
      \caption{Illustration of the rate-aware adaptive temporal prediction method.}
   \label{fig:raatp}
\end{figure}
%-------------------------------------------------------------------------
Guided by this principle, as shown in Fig.~\ref{fig:raatp}, we design a multi‑branch context prediction architecture that produces a candidate set \(\mathcal{C} = \{C_t^0, C_t^1, C_t^2\}\) through structurally diverse processing:
\begin{itemize}
\item Full‑resolution branch \(C_t^0\):
\begin{equation}
C_t^0 = \phi_0(\hat{F}_{t-1}; \theta_0).
\end{equation}
This branch operates at the original feature resolution, preserving the fine‑grained spatial details present in the reference feature \(\hat{F}_{t-1}\).

\item Low‑resolution parallel branches \(C_t^1, C_t^2\):
\begin{equation}
\begin{aligned}
&C_t^1 = \text{Up}(\phi_1(\text{Down}(\hat{F}_{t-1}); \theta_1)), \\
&C_t^2 = \text{Up}(\phi_2(\text{Down}(\hat{F}_{t-1}); \theta_2)),
\end{aligned}
\end{equation}
where \(\text{Down}(\cdot)\) denotes a bilinear down‑sampling operation and \(\text{Up}(\cdot)\) denotes a bilinear up‑sampling operation.
The down‑sampling operation reduces the spatial resolution, which enlarges the effective receptive field of subsequent convolutions and discards high‑frequency information. The two branches receive the same down‑sampled input but are initialized and trained with independent weights \(\theta_1\) and \(\theta_2\), allowing them to learn complementary features and produce distinct prediction candidates.
\end{itemize}
This heterogeneous design promotes a larger ensemble diversity term, providing a joint estimate against complex temporal misalignment. The structural diversity across resolutions and parallel streams enables the ensemble to capture a range of possible motion patterns.

To perform rate‑distortion optimized mode selection without discrete search, we let the learnable quantization vectors
\(q_c^0, q_c^1, q_c^2 \in \mathbb{R}^{C \times 1 \times 1}\), which are part of the module bank and indexed by QP, serve as fusion weights.
The final temporal context \(C_t\) is obtained as:
\begin{equation}
C_t = f_{\text{fuse}}\Big( \big(q_c^0 \odot C_t^0\big) \;\|\; \big(q_c^1 \odot C_t^1\big) \;\|\; \big(q_c^2 \odot C_t^2\big) \Big),
\end{equation}
where \(\odot\) is channel‑wise multiplication, \(\|\) denotes concatenation along the channel dimension, and \(f_{\text{fuse}}\) is a subsequent depth‑wise convolutional fusion layer.
These \(q_c^k\) function as differentiable proxy variables~\cite{jia2025towards} that directly modulate the contribution of each branch in the feature domain, driven by the rate‑distortion optimization. 
The role of \(q_c^k\) can be understood by examining the gradient of the rate-distortion objective with respect to these quantization vectors. Since \(q_c^k\) multiplies the branch output \(C_t^k\) before the fusion network, the chain rule gives
\begin{equation}
\frac{\partial \mathcal{L}}{\partial q_c^k}
= \left( \lambda \frac{\partial d}{\partial (q_c^k \odot C_t^k)} + \frac{\partial R}{\partial (q_c^k \odot C_t^k)} \right) \odot C_t^k,
\label{eq:grad_q}
\end{equation}
where \(\odot\) denotes element-wise multiplication.
This gradient jointly combines a distortion-driven component and a rate-driven component, both of which are balanced by the Lagrange multiplier \(\lambda\).
When \(\lambda\) is small and the rate penalty dominates, the optimization tends to suppress high‑frequency channels in the full‑resolution branch \(C_t^0\), allowing the coarser branches \(C_t^1\) and \(C_t^2\) to contribute more to the fused context \(C_t\).
When \(\lambda\) is large and the distortion penalty dominates, the suppression is relaxed, and the full‑resolution branch can contribute its fine‑grained details more fully.
This rate‑dependent adaptive selection emerges from end‑to‑end training, thereby restoring the adaptivity of temporal prediction to motion complexity and bitrate constraints that was lost in deterministic implicit prediction.

\subsection{Decomposition-based Spatial Rate Control}\label{sec:spatial_rate_control}
Existing neural video codecs employ a two‑level quantization mechanism~\cite{li2023neural,li2024neural,jia2025towards}. The first level operates in the encoder and decoder: learnable one‑dimensional quantization vectors \(q^E, q^D \in \mathbb{R}^{C \times 1 \times 1}\) uniformly modulate the encoder intermediate feature \(F_t^E\) and the decoder intermediate feature \(F_t^D\) across all spatial locations:
\begin{equation}
\tilde{F}_t^E = F_t^E \odot q^E, \quad \tilde{F}_t^D = F_t^D \odot q^D.
\label{eq:uniform_quant}
\end{equation}
The second level is realized by the entropy model: in addition to predicting the mean \(\mu_t\) and scale \(\sigma_t\), the entropy model learns a quantization step \(Q_t\) of the same size as the latent representation \(y_t\). Before rounding ($\left\lfloor \right\rceil$) at the encoder, \(y_t\) is divided by \(Q_t\) element‑wise; at the decoder, \(\hat{y}_t\) is multiplied back by \(Q_t\):
\begin{equation}
y_t^{\text{quant}} = \left\lfloor \frac{y_t}{Q_t} \right\rceil, \quad \hat{y}_t = y_t^{\text{quant}} \odot Q_t.
\label{eq:entropy_quant}
\end{equation}

Since the first‑level quantization is spatially uniform, and although the second‑level quantization enables some degree of content‑adaptive spatial bit allocation, it operates at the low spatial resolution of \(y_t\) (16\(\times\) smaller than the original frame), which limits its allocation precision. More importantly, once trained, users cannot freely control the quantization strength across spatial regions at test time. \par
%-------------------------------------------------------------------------
\begin{figure}[t]
  \centering
   \includegraphics[width=0.65\linewidth]{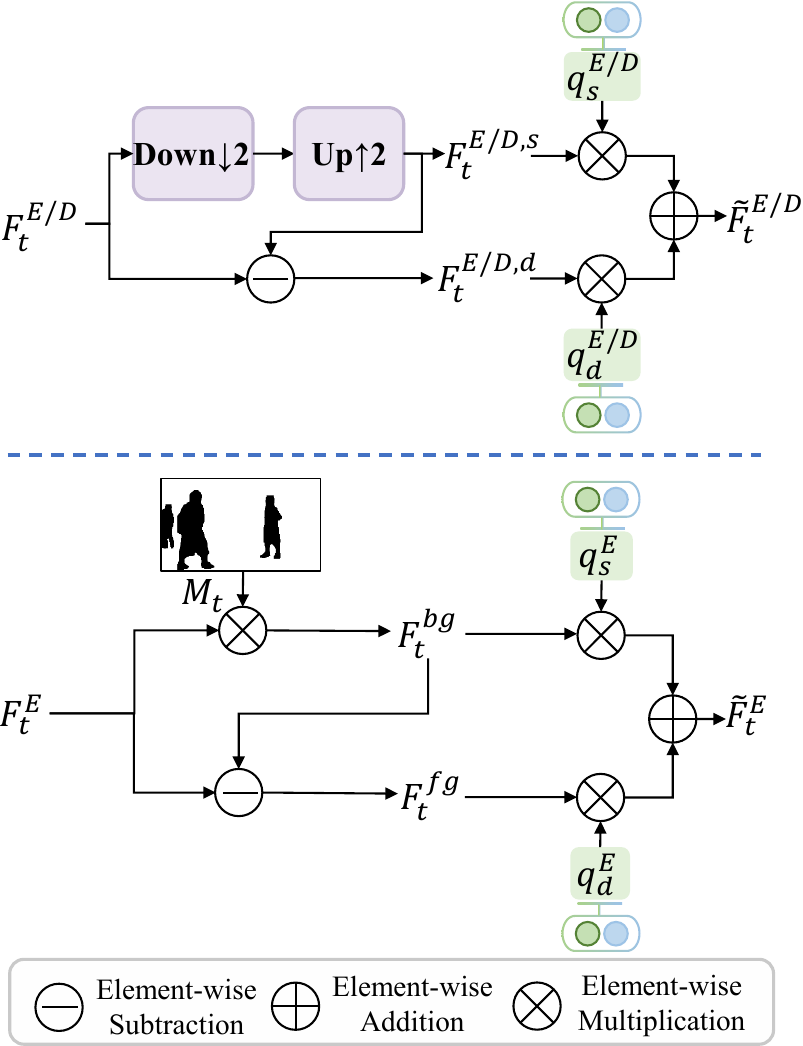}
      \caption{Illustration of the decomposition-based spatial rate control method.}
   \label{fig:src}
\end{figure}
%-------------------------------------------------------------------------

To achieve direct spatial rate control at test time, we propose a spatial decomposition‑based quantization mechanism. For the encoder intermediate feature \(F_t^E\) and the decoder intermediate feature \(F_t^D\), we obtain their smooth structural component by bilinearly down‑sampling the features to half resolution and then up‑sampling it back to the original resolution:
\begin{equation}
F_t^{E/D,s} = \text{Up}\bigl(\text{Down}(F_t^{E/D})\bigr).
\label{eq:spatial_dec}
\end{equation}
The detail component captures the texture and edge information filtered out by down‑sampling, and is computed as the residual:
\begin{equation}
F_t^{E/D,d} = F_t^{E/D} - F_t^{E/D,s}.
\label{eq:spatial_res}
\end{equation}
Each component is then modulated by its own learnable quantization vector \(q_{s}^{E/D}, q_{d}^{E/D} \in \mathbb{R}^{C \times 1 \times 1}\), and the two are summed back together:
\begin{equation}
\tilde{F}_t^{E/D} = F_t^{E/D,s} \odot q_{s}^{E/D} + F_t^{E/D,d} \odot q_{d}^{E/D}.
\label{eq:spatial_quant}
\end{equation}
The learnable quantization vectors \(q_{\mathrm{s}}^{E/D}\) and \(q_{\mathrm{d}}^{E/D}\) are part of the module bank and are indexed by the current QP. Trained end‑to‑end under the rate‑distortion loss, they allow the network to allocate bits unevenly across spatial structures. The structural component can be heavily quantized to save rate, while the detail component can retain finer precision.

A crucial property of this design is that the down‑up sampling-based decomposition in the encoder can be replaced at test time by an arbitrary spatial mask-based decomposition. Concretely, let \(M_t \in \{0,1\}^{1 \times H \times W}\) be a user‑defined binary ROI mask, down‑sampled to the resolution of \(F_t^E\). The mask‑based decomposition can be performed as:
\begin{equation}
F_t^{bg} = F_t^E \odot M_t, \qquad F_t^{fg} = F_t^E \odot (1 - M_t).
\label{eq:roi_split}
\end{equation}
Then we apply the structural quantization vector \(q_{s}^E\) to the background and the detail quantization vector \(q_{d}^E\) to the foreground:
\begin{equation}
\tilde{F}_t^E = F_t^{bg} \odot q_{s}^E + F_t^{fg} \odot q_{d}^E.
\label{eq:roi_quant}
\end{equation}
Since \(q_{d}^E\) has learned to preserve fine‑grained information and \(q_{s}^E\) has learned to encode smooth content at lower fidelity during training, this substitution naturally grants higher reconstruction quality to the user‑specified foreground regions while compressing the background regions more efficiently. The decoder continues to use the original down‑up decomposition, so no modification is needed at the receiver side. The mask \(M_t\) itself is not transmitted, incurring zero bitrate overhead.

\subsection{Perceptual Switching}\label{sec:perceptual_switching}
The perception‑distortion trade‑off~\cite{blau2018perception} shows that the fidelity‑optimal and perception‑optimal reconstructions occupy distinct points on the rate‑distortion‑perception surface, and therefore require different sets of model parameters. Following this, existing neural video codecs train two separate models. A model optimized for signal fidelity outputs the conditional expectation \(\mathbb{E}[X|Y]\), where \(X\) denotes the original signal and \(Y\) denotes the available observations at the decoder, minimizing pixel‑level error but producing smooth reconstructions.  A model optimized for perceptual quality sacrifices pixel fidelity to produce visually plausible textures and achieve better perceptual quality. However, this two‑model solution substantially increases storage and maintenance costs, which is particularly burdensome for resource‑constrained applications.

A closer examination of the theory, however, reveals an unexploited degree of freedom. The perception‑distortion trade‑off constrains only the conditional distribution \(p_{\hat{X}|Y}\) that the codec realizes, not the network architecture that produces it. In other words, while the two modes must correspond to different conditional distributions, nothing in the theory requires them to be realized by two entirely separate networks. This observation motivates a design in which the model backbone is shared between the two modes, and only a small set of mode‑specific parameters is switched. Building on this insight, we propose a perceptual switching method.

We first identify the quantization vectors as the primary mode‑specific parameters. After end‑to‑end training for signal fidelity, the backbone convolutional layers have learned to extract generic feature representations from the input frames. The learnable quantization vectors apply channel‑wise modulation to these features. Different quantization vectors do not alter how the backbone convolution kernels extract features, but only change the retained amplitude of each channel. Hence, the backbone convolution kernels can be shared, while the quantization vectors can be switched. This design can be further understood from the perspective of covariance. Consider a feature representation \(F \in \mathbb{R}^{C \times H \times W}\) before modulation, and let \(\tilde{F} = F \odot q\) be the feature after modulation by a quantization vector \(q \in \mathbb{R}^{C \times 1 \times 1}\). The channel covariance matrix \(\Sigma_F = \mathbb{E}[(F - \mu_F)(F - \mu_F)^\top]\) becomes:
\begin{equation}
\Sigma_{\tilde{F}} = \operatorname{diag}(q) \; \Sigma_F \; \operatorname{diag}(q)^{\top},
\label{eq:cov_rescale_percep}
\end{equation}
where \(\operatorname{diag}(q)\) is a diagonal matrix with the entries of \(q\) on the diagonal, and \(\mu_F\) is the channel‑wise mean of \(F\). This identity shows that replacing the quantization vectors is equivalent to independently rescaling the diagonal entries of the feature covariance matrix. Two different sets of quantization vectors therefore produce different feature energy distributions, each corresponding to one of the two modes.\par
%-------------------------------------------------------------------------
\begin{figure}[t]
  \centering
   \includegraphics[width=0.6\linewidth]{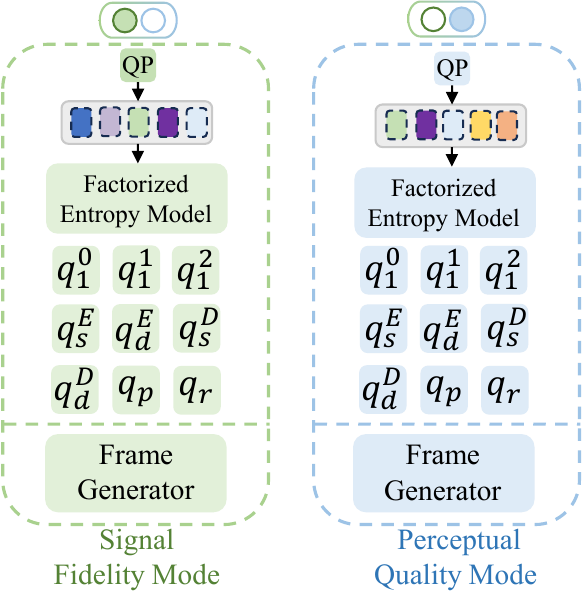}
      \caption{Illustration of the perceptual switching method.}
   \label{fig:ps}
\end{figure}
%-------------------------------------------------------------------------
Replacing the quantization vectors leads to two further requirements. First, the factorized entropy model must be updated synchronously. When the distribution of the hyper‑prior latent \(\hat{z}_t\) shifts due to the changed quantization vectors, the original entropy model would produce inaccurate probability estimates. Hence, the two modes require separate factorized entropy models \(\phi_z^{\text{Fid}}\) and \(\phi_z^{\text{Per}}\). Second, the frame generator \(\theta_r\), which maps the decoded feature \(\hat{F}_t\) to the reconstructed frame \(\hat{x}_t\), also needs to be replaced. When the statistics of the decoded feature change under perceptual quantization vectors, the original frame generator may no longer produce outputs aligned with perceptual objectives. The two modes therefore also require separate frame generators \(\theta_r^{\text{Fid}}\) and \(\theta_r^{\text{Per}}\).

Building on the above analysis, as presented in Fig.~\ref{fig:ps}, we maintain two module banks (\(\mathcal{B}^{\text{Fid}}\), \(\mathcal{B}^{\text{Per}}\)) and two frame generators (\(\theta_r^{\text{Fid}}, \theta_r^{\text{Per}}\)) for the two modes, both having the same structure. Each module bank is indexed by QP and contains a set of channel‑wise quantization vectors---the structural/detail quantization vectors \(q_s^{E/D}, q_d^{E/D}\), the temporal prior quantization vector \(q_p\), the context fusion quantization vectors \(q_c^0, q_c^1, q_c^2\), the reconstruction quantization vector \(q_r\), and a factorized entropy model \(\phi_z\). 

Training proceeds in two stages. In the first stage, all convolutional layers in the contextual encoder, contextual decoder, and temporal prediction module, together with \(\mathcal{B}^{\text{Fid}}\) and the frame generator \(\theta_r^{\text{Fid}}\), are jointly trained under a rate‑fidelity loss. In the second stage, all convolutional weights are frozen. \(\mathcal{B}^{\text{Per}}\) and the frame generator \(\theta_r^{\text{Per}}\) are initialized from \(\mathcal{B}^{\text{Fid}}\) and \(\theta_r^{\text{Fid}}\), respectively, and fine‑tuned using a perceptual loss. Because the backbone is trained once for signal fidelity and then frozen, the fidelity mode remains completely unaffected when the perceptual mode is added.
At inference time, users can switch between two modes by changing the loaded module bank and frame generator parameters, incurring negligible additional storage and identical computational cost.

\section{Experiments}\label{sec:experiment}
\subsection{Datasets}
We train our model on the training split of the Vimeo-90k dataset~\cite{xue2019video}. Following~\cite{jia2025towards,sheng2025prediction}, we further process the original Vimeo video clips into longer sequences for cascaded training. For evaluation, we use three widely adopted benchmark datasets. The HEVC Common Test Sequences~\cite{sullivan2012overview} (Classes B, C, D, and E) cover resolutions from 416$\times$240 to 1920$\times$1080, with content ranging from static conference scenes to fast‑paced sports. The UVG dataset~\cite{mercat2020uvg} contains seven 1080p high‑frame‑rate sequences. The MCL‑JCV dataset~\cite{wang2016mcl} consists of thirty 1080p sequences with diverse content. All test sequences are in YUV420 format.
\subsection{Test Details}
For the traditional codec, we compare with the official reference software VTM‑17.0 of the H.266/VVC~\cite{bross2021overview} standard under the \emph{encoder\_lowdelay\_vtm.cfg} configuration. 
%The detailed encoding commands for VTM-17.0 are shown as follows.
%\begin{itemize}
%    \item -c $\{\emph{config file name}\}$  \mbox{-}\mbox{-}InputFile=$\{\emph{input video name}\}$  \mbox{-}\mbox{-}FrameRate=$\{\emph{frame rate}\}$     \mbox{-}\mbox{-}DecodingRefreshType=2  \mbox{-}\mbox{-}InputBitDepth=8 \mbox{-}\mbox{-}OutputBitDepth=8 \mbox{-}\mbox{-}OutputBitDepthC=8 \mbox{-}\mbox{-}SourceWidth=$\{\emph{width}\}$  \mbox{-}\mbox{-}SourceHeight=$\{\emph{height}\}$ 
%\mbox{-}\mbox{-}IntraPeriod=$\{\emph{intra period}\}$ 
% \mbox{-}\mbox{-}FramesToBeEncoded=$\{\emph{frames}\}$ \mbox{-}\mbox{-}QP=$\{\emph{qp}\}$ 
% \mbox{-}\mbox{-}Level=6.2 \mbox{-}\mbox{-}BitstreamFile=$\{\emph{bitstream file name}\}$
%\end{itemize}
For neural codecs, we compare with DCVC‑DC~\cite{li2023neural}, DCVC‑FM~\cite{li2024neural}, PRAVC~\cite{sheng2025prediction}, and the state‑of‑the‑art real‑time codec DCVC‑RT~\cite{jia2025towards}.  We also include our reproduced version of DCVC‑RT, denoted as DCVC‑RT$^*$, which serves as the direct baseline of our model. Due to limited GPU memory, we are unable to perform the cascaded training~\cite{sheng2022temporal} with 64 frames at a resolution of 512$\times$512 as done in the original DCVC‑RT. Consequently, DCVC‑RT$^*$ achieves lower compression performance than the released model.\par
%-------------------------------------------------------------------------
\begin{figure*}[t]
  \centering
   \includegraphics[width=0.9\linewidth]{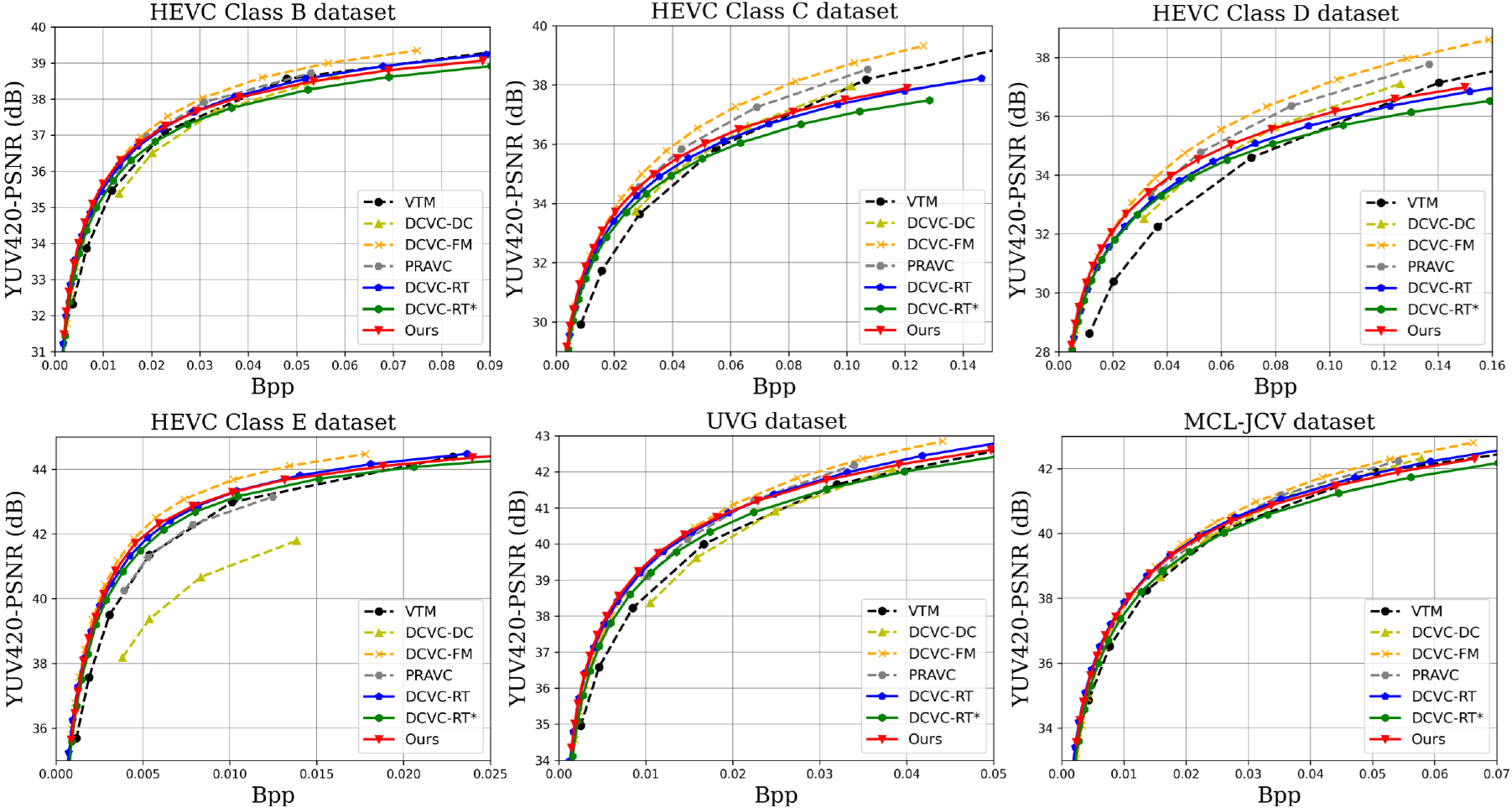}
      \caption{Rate‑distortion curves on HEVC, UVG, and MCL‑JCV datasets. All frames are coded with intra period --1 in YUV420 color space. The quality metric is the weighted YUV PSNR where the Y, U, and V components are combined with a 6:1:1 ratio.}
   \label{fig:rd_curves}
\end{figure*}
%-------------------------------------------------------------------------
%------------------------------------------------------------------------
\begin{table*}[t]
\caption{BD-rate(\%) comparison in YUV420 colorspace measured with PSNR. The anchor is VTM-17.0. All frames are tested with the intra period set to --1.}
  \centering
\scalebox{1}{
\begin{threeparttable}
\begin{tabular}{l|c|c|c|c|c|c|c}
\toprule[1.5pt]
               & HEVC Class B  & HEVC Class C  &HEVC Class D &HEVC Class E &UVG            &MCL-JCV  &Average\\ \hline
VTM            &0.0            &0.0            &0.0          &0.0          &0.0            &0.0      &0.0\\ \hline
DCVC-DC        &13.4           &--4.5          &--16.8       &88.8         &7.8            &--2.0    &14.5\\ \hline
DCVC-FM        &--18.0         &--30.7         &--39.1       &--29.6       &--21.7         &--12.1   &--25.2 \\ \hline
PRAVC          &--9.5          &--18.0         &--27.0       &3.7          &--15.7         &--7.3    &--12.3\\ \hline
DCVC-RT        &--13.6         &--17.4         &--25.7       &--21.9       &--22.2         &--12.1   &--18.8 \\ \hline
DCVC-RT*       &--3.8          &--12.9         &--25.5       &--14.4       &--12.0         &--0.5    &--11.5\\ \hline
Ours           &--16.3         &--25.4         &--35.8       &--22.3       &--24.1         &--10.6   &--22.4 \\ 
\bottomrule[1.5pt]
\end{tabular}
  \begin{tablenotes}
   \item \footnotesize \dag DCVC‑RT$^*$ is our reproduced version of DCVC‑RT, which serves as the direct baseline of our model. Due to limited GPU memory, we are unable to perform the cascaded training with 64 frames at a resolution of 512$\times$512 as done in the original DCVC‑RT. Consequently, DCVC‑RT$^*$ achieves  lower compression performance than the released model.
  \end{tablenotes}
\end{threeparttable}
}
\label{table:ip1_yuv420_psnr}
\end{table*}
%-------------------------------------------------------------------------
%-------------------------------------------------------------------------
\begin{figure}[!htb]
  \centering
   \includegraphics[width=0.7\linewidth]{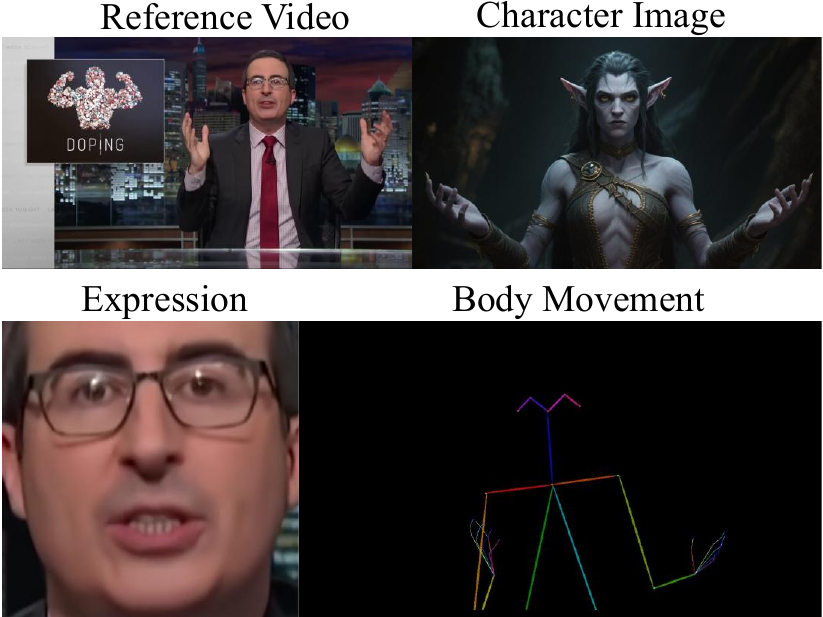}
      \caption{Illustration of video motion control as a typical ROI task. The generative model only extracts the character's facial expressions and body movements from the reference video, while the background is largely ignored.}
   \label{fig:ROI2}
\end{figure}
%-------------------------------------------------------------------------

%-------------------------------------------------------------------------
\begin{figure*}[!htb]
  \centering
   \includegraphics[width=0.67\linewidth]{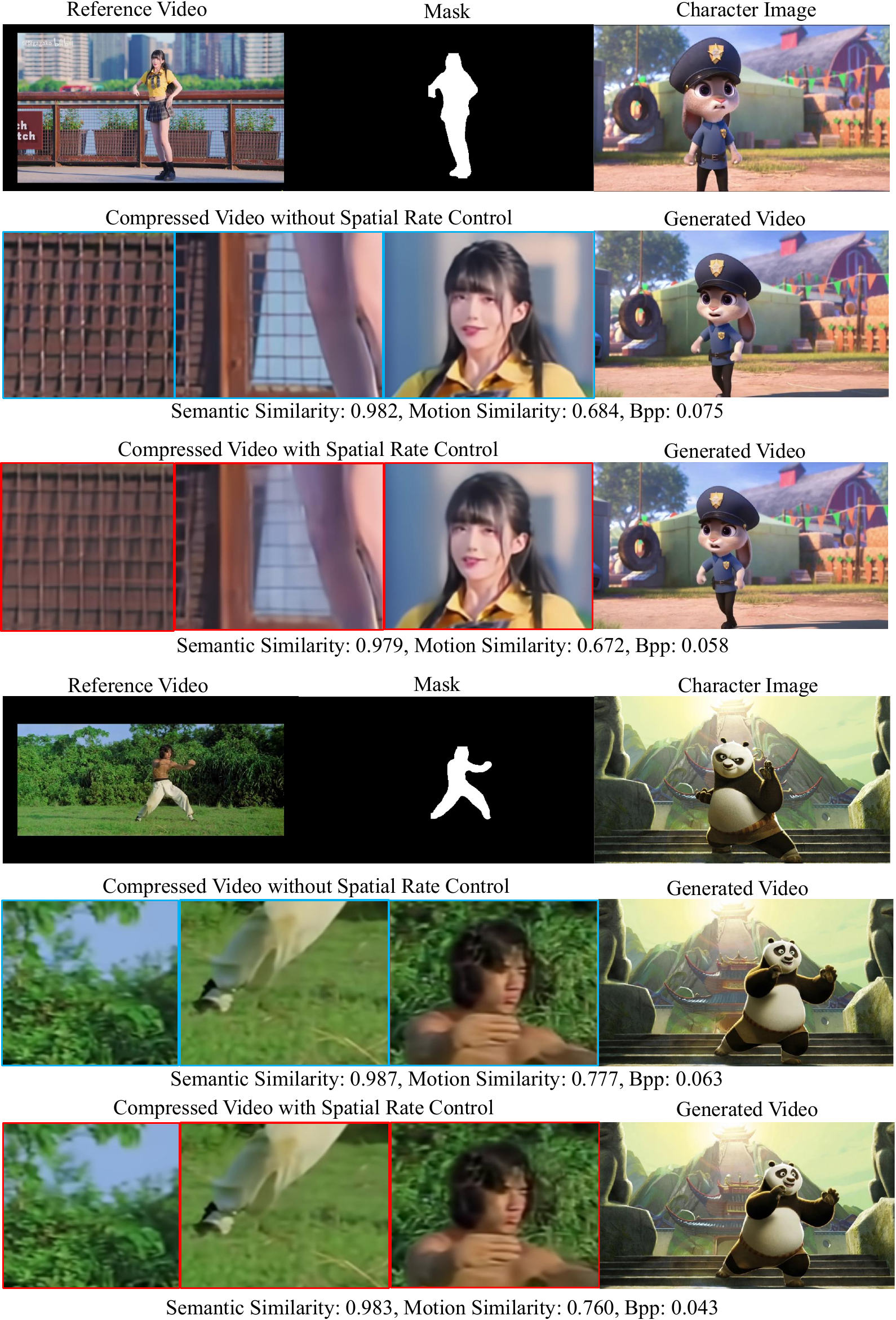}
      \caption{Comparison of reference video compression and generated animation with/without our spatial rate control on the \textit{Dancer} and \textit{KungFu} sequences.}
   \label{fig:ROI1}
\end{figure*}
%-------------------------------------------------------------------------

We conduct three sets of experiments to evaluate our codec from different perspectives. For signal fidelity, we assess rate‑distortion performance using YUV420 PSNR with VTM‑17.0 as the anchor. For perceptual quality, we evaluate DISTS~\cite{ding2020image} and LPIPS~\cite{zhang2018unreasonable} with DCVC‑FM as the anchor. For spatial rate control, we test on two representative videos---a dancing sequence and a martial‑arts sequence---using a CLIP~\cite{radford2021learning} semantic similarity and a motion similarity as evaluation metrics. All frames are encoded with intra period set to $-1$ and reported in the YUV420 color space. 
%The bitrate is measured in bits per pixel (bpp), calculated as the total number of bits divided by the total number of pixels across all frames. Rate‑distortion performance is assessed by the Bj{\o}ntegaard Delta bit‑rate (BD‑Rate)~\cite{bjontegaard2001calculation}.
\subsection{Training Details}
To support variable‑rate coding within a single model, we randomly sample a QP from the range $[0,63]$ in each training iteration.
To implement a hierarchical quality structure, we adopt the QP offset pattern $[0,8,0,4,0,4,0,4]$ for a group of 8 pictures and assign distortion weights of [$1.2$, $0.5$, $0.9$, $0.5$] to the four corresponding hierarchical levels.
Each QP serves as an index to select a corresponding set of components in the module bank.
The Lagrangian multiplier $\lambda$ is interpolated between $1$ and $768$ according to the sampled QP.
The loss function differs between the fidelity mode and perceptual mode. For signal fidelity training, we use a scaled YUV mean squared error (MSE) loss with the Y, U, and V components weighted at a ratio of $6{:}1{:}1$. For perceptual quality training, the distortion term is a weighted combination of YUV MSE, LPIPS, and a GAN loss~\cite{yang2022perceptual}.
The model is optimized with the AdamW optimizer~\cite{kingma2014adam} using a batch size of $8$.
All training is performed on $8$ NVIDIA RTX 3090 GPUs.

\subsection{Experimental Results for Signal Fidelity}
Table~\ref{table:ip1_yuv420_psnr} reports the BD-rate savings in YUV420 PSNR with VTM‑17.0 as the anchor. 
Our codec achieves an average BD-rate of $-22.4\%$. 
When compared with codecs whose computational complexity exceeds millions of MACs per pixel, our codec substantially outperforms DCVC‑DC ($14.5\%$) and PRAVC ($-12.3\%$), and remains slightly below DCVC‑FM ($-25.2\%$) in overall compression efficiency. When compared with lightweight real‑time codecs, our codec delivers over $10\%$ bitrate saving over our reproduced baseline DCVC‑RT$^*$ ($-11.5\%$) under the same constrained training protocol.
This substantial improvement directly validates the effectiveness of our proposed rate-aware adaptive temporal prediction and finer‑grained spatial bit allocation.
More importantly, our codec even surpasses the officially released DCVC‑RT ($-22.4\%$ vs. $-18.8\%$) on average.
This improvement is consistent across most individual datasets, with notable gains on HEVC Class~C ($-25.4\%$ vs. $-17.4\%$) and HEVC Class~D ($-35.8\%$ vs. $-25.7\%$). Nevertheless, on high-resolution sequences such as 1080p videos, the rate-distortion curves in Fig.~\ref{fig:rd_curves} show that our codec performs slightly below the released DCVC‑RT at high bitrates.
This is mainly due to our limited GPU memory, which prevents us from using the same training protocol of the original DCVC‑RT. We can only train with smaller-resolution crops of $256\times384$ (instead of $512\times512$), and although we also use $64$-frame sequences, we are forced to adopt a grouped cascaded training scheme~\cite{jiang2025ecvc} with detached computation graphs rather than the unrolled continuous cascading of the official implementation.
These constraints slightly reduce the model's capacity at very high bitrates. We believe that with access to GPUs with larger memory, our model would achieve further performance gains.

\subsection{Experimental Results for Spatial Rate Control}
We evaluate our spatial rate control method on a video motion control task, a representative ROI coding scenario. In this task, a user uploads a reference video and a character image to a cloud server, and a generative model animates the character to mimic the motion of the reference video. As exemplified by recent project such as Seedance~\cite{seedance2026seedance}, the generative model only needs to extract the character's facial expressions and body movements from the reference video, while the background is largely ignored, as illustrated in Fig.~\ref{fig:ROI2}. Consequently, the foreground (the person) requires high reconstruction quality, while the background can be heavily compressed without affecting the generated video. This makes the problem a natural fit for ROI coding.

We test on two representative reference videos, including a dancing video (\textit{Dancer}) and a martial‑arts video (\textit{KungFu}). Fig.~\ref{fig:ROI1} compares the reconstructed reference videos with and without our spatial rate control. As expected, the background regions---e.g., the fence in \textit{Dancer} and the grass in \textit{KungFu}---become noticeably blurred when spatial rate control is applied, because they are quantized more heavily. In contrast, the foreground remains sharp. We feed the compressed reference videos into the same video motion control model~\cite{seedance2026seedance}. Visual inspection shows that the generated videos are virtually indistinguishable from each other.

%-------------------------------------------------------------------------
\begin{figure*}[!htb]
  \centering
   \includegraphics[width=0.85\linewidth]{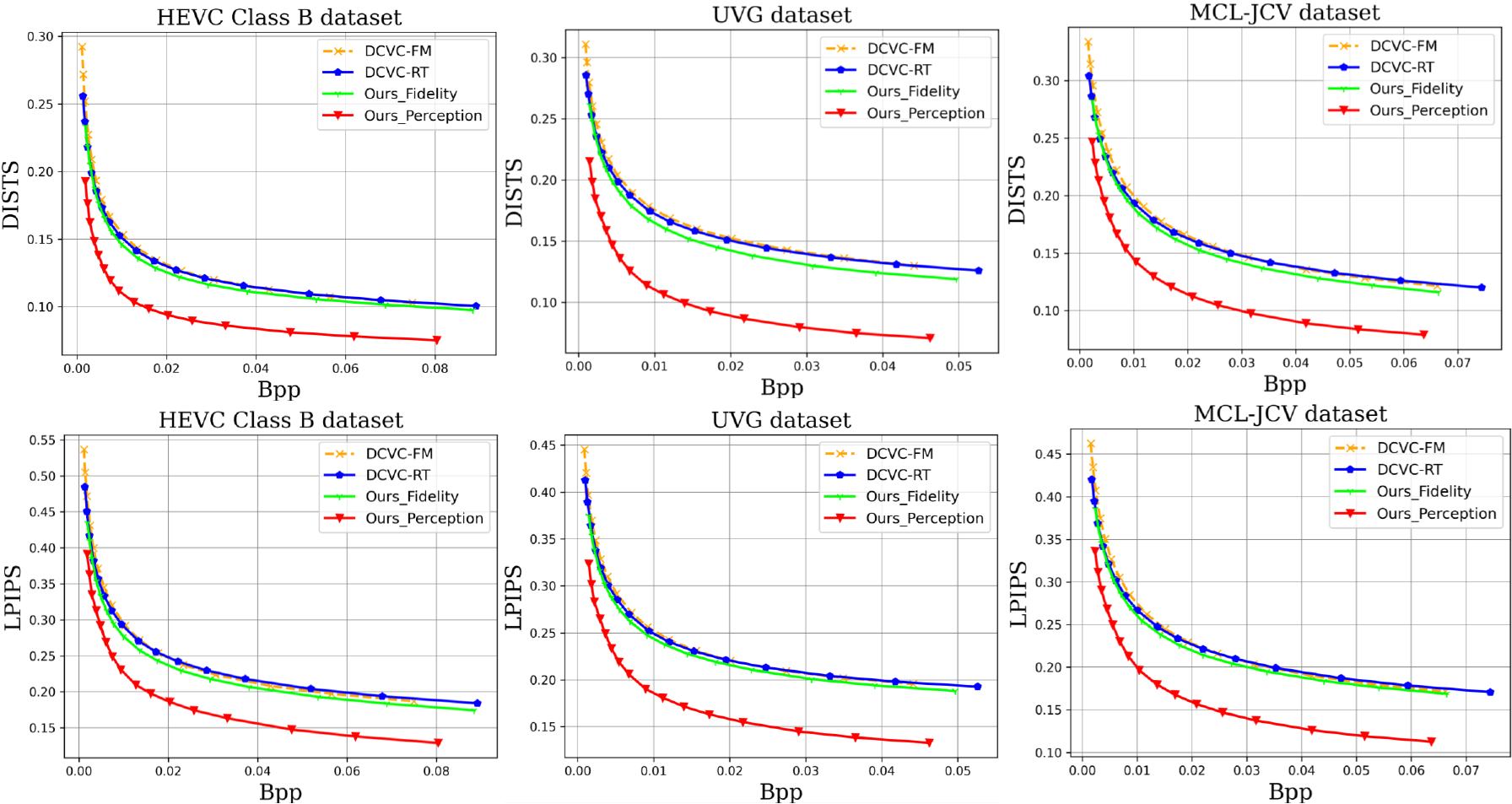}
      \caption{Rate‑perception curves on HEVC, UVG, and MCL‑JCV datasets. All frames are coded with intra period --1 in YUV420 color space. The reconstructed videos are converted to RGB before computing DISTS and LPIPS.}
   \label{fig:GAN_RD}
\end{figure*}
%-------------------------------------------------------------------------

\begin{table*}[htb]
\caption{BD-rate(\%) comparison in terms of DISTS (left) and LPIPS (right).}
\centering
\scalebox{0.9}{
\begin{threeparttable}
\footnotesize
\begin{tabular}{l|c|c|c|c|c|c|c}
\toprule[1.5pt]
               & HEVC Class B & HEVC Class C & HEVC Class D & HEVC Class E & UVG      & MCL-JCV & Average \\ \hline
DCVC-FM        & 0.0 / 0.0  & 0.0 / 0.0  & 0.0 / 0.0  & 0.0 / 0.0  & 0.0 / 0.0 & 0.0 / 0.0 & 0.0 / 0.0 \\ \hline
DCVC-RT        & --11.1 / --8.1 & 0.3 / --0.3 & --0.4 / --1.7 & --7.5 / --0.2 & --12.1 / --8.8 & --11.2 / --10.5 & --7.0 / --4.9 \\ \hline
Ours (Fidelity)& --17.4 / --20.3 & --4.3 / --11.1 & --6.7 / --12.5 & --8.1 / --0.4 & --23.7 / --16.8 & --16.9 / --16.2 & --12.9 / --12.9 \\ \hline
Ours (Perception)& --69.0 / --57.7 & --46.1 / --37.0 & --34.2 / --31.7 & --56.7 / --36.5 & --75.7 / --66.8 & --62.0 / --62.6 & --57.3 / --48.7 \\
\bottomrule[1.5pt]
\end{tabular}
\end{threeparttable}
}
\label{table:ip1_combined_dists_lpips}
\end{table*}
%-------------------------------------------------------------------------
\begin{figure*}[t]
  \centering
   \includegraphics[width=0.82\linewidth]{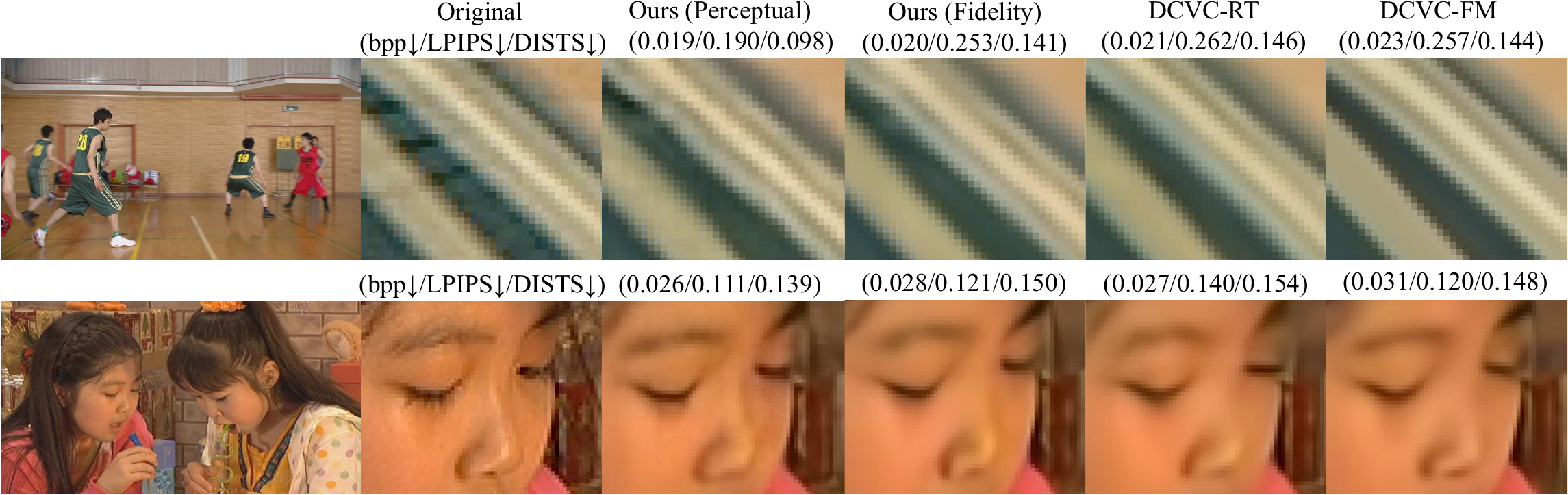}
      \caption{Visual comparison between DCVC‑FM~\cite{li2024neural},  DCVC‑RT~\cite{jia2025towards}, our codec in signal fidelity mode, and our codec in perceptual mode.}
   \label{fig:GAN_subjective}
\end{figure*}
%-------------------------------------------------------------------------
For objective evaluation, we compute two similarity metrics between the video frames (denoted as \(\{G_i^o\}_{i=1}^{N}\)) generated from the original uncompressed reference video and those (denoted as \(\{G_i^c\}_{i=1}^{N}\)) generated from the compressed reference video. The first metric is CLIP semantic similarity. Let \(f_{\text{CLIP}}(\cdot)\) denote the CLIP image encoder (ViT-base-patch32) that maps an frame to a normalized feature vector. The semantic similarity for frame \(i\) is the cosine similarity between \(f_{\text{CLIP}}(G_i^o)\) and \(f_{\text{CLIP}}(G_i^c)\). The average over all frames is:
\begin{equation}
\text{CLIP-Sim} = \frac{1}{N} \sum_{i=1}^{N} \frac{f_{\text{CLIP}}(G_i^o) \cdot f_{\text{CLIP}}(G_i^c)}{\|f_{\text{CLIP}}(G_i^o)\| \|f_{\text{CLIP}}(G_i^c)\|}.
\end{equation}
The second metric is motion similarity. Let \(\text{flow}(\cdot,\cdot)\) denote the dense optical flow field computed between two frames. For adjacent frames \(i\) and \(i+1\), define \(\mathbf{u}_i^{G^c} = \text{flow}(G_i^c, G_{i+1}^c)\) and \(\mathbf{u}_i^{G^o} = \text{flow}(G_i^o, G_{i+1}^o)\). The motion similarity is the average cosine similarity over all adjacent frame pairs:
\begin{equation}
\text{Motion-Sim} = \frac{1}{N-1} \sum_{i=1}^{N-1} \frac{\mathbf{u}_i^{G^c} \cdot \mathbf{u}_i^{G^o}}{\|\mathbf{u}_i^{G^c}\| \|\mathbf{u}_i^{G^o}\|}.
\end{equation}
As shown in Fig.~\ref{fig:ROI1}, both metrics remain nearly identical for the videos generated with and without spatial rate control, while our method achieves 20--30\% bitrate savings on the reference video. For the \textit{KungFu} sequence, the CLIP similarity is 0.987 vs.\ 0.983, and the flow similarity is 0.777 vs.\ 0.760, while the bitrate is reduced by 31.7\%. Similar results are observed on the \textit{Dancer} sequence. These results confirm that our spatial rate control can effectively allocate bits to semantically relevant regions without harming the downstream generative task.

\subsection{Experimental Results for Perceptual Quality}
We present the rate-perception curves in terms of DISTS and LPIPS in Fig.~\ref{fig:GAN_RD}. Detailed BD‑rate saving values are reported in Tables~\ref{table:ip1_combined_dists_lpips}, with DCVC‑FM as the anchor. The results show that our codec with the perceptual mode achieves an average BD‑rate of $-57.3\%$ in DISTS and $-48.7\%$ in LPIPS, substantially outperforming both DCVC‑FM and the released DCVC‑RT ($-7.0\%$ DISTS, $-4.9\%$ LPIPS). Fig.~\ref{fig:GAN_subjective} provides visual comparisons on two sequences. On \emph{BasketballDrive\_1920x1080\_50}, our perceptual mode operates at a lower bitrate of 0.019\,bpp, yet achieves markedly better DISTS of 0.098 and LPIPS of 0.190, compared with 0.144 and 0.257 for DCVC‑FM at 0.023\,bpp. The visual gain is evident on the player's shorts, where the stripe patterns are noticeably sharper and more faithfully rendered. On \emph{BlowingBubbles\_416x240\_50}, our codec delivers a DISTS of 0.139 and an LPIPS of 0.111 at 0.026\,bpp, while DCVC‑FM obtains 0.148 and 0.120 at 0.031\,bpp. The improvement is visible on the girl's eyebrows, which appear clearer and more naturally textured. These results confirm that our proposed perceptual switching method yields substantial perceptual gains.

%-------------------------------------------------------------------------
\begin{table}[t]
 \centering
 \caption{Encoding/decoding time and model complexity comparison for one 1080p video frame.}
\scalebox{1}{
\begin{tabular}{c|c|c|c|c}
\toprule[1.5pt]
Schemes  & Encoding & Decoding& MACs/pixel & Params          \\ \hline
DCVC-DC     & 1.21\,s& 0.64\,s  & 1397.90K&18.45M \\ \hline
DCVC-FM     & 0.74\,s& 0.53\,s  & 1180.77K &17.02M \\ \hline
PRAVC       & 1.11\,s& 0.85\,s  & 1963.56K& 19.28M  \\ \hline
DCVC-RT     & 0.020\,s& 0.021\,s  & 195.23K& 19.93M  \\ \hline
Ours        & 0.027\,s& 0.028\,s  & 223.00K& 24.89M  \\
\bottomrule[1.5pt]
\end{tabular}}
\label{time}
\end{table}
%-------------------------------------------------------------------------

\subsection{Complexity Analysis}
Table~\ref{time} compares the encoding and decoding speed, MACs per pixel, and model parameters of our codec with those of several representative neural video codecs. All measurements are taken at 1080p resolution on a single NVIDIA RTX 3090 GPU. The results show that our codec encodes one 1080p frame in $0.027$ seconds and decodes in $0.028$ seconds, slightly higher than the real‑time baseline DCVC‑RT ($0.020$\,s and $0.021$\,s). The computational cost increases from $195.23$K to $223.00$K\,MACs per pixel, and the model size grows from $19.93$M to $24.89$M parameters. Despite these increases, our codec remains firmly in the real‑time regime. Compared with the non‑real‑time codecs DCVC‑DC, DCVC‑FM, and PRAVC, it is more than $20\times$ faster while delivering competitive or superior compression performance. The modest additional complexity over DCVC‑RT is the cost paid for the temporal, spatial, and perceptual adaptivity that no existing real‑time neural codec provides.

%-------------------------------------------------------------------------
 \begin{table}[t]
\caption{Effectiveness of proposed rate-aware adaptive temporal prediction (RAATP) method and decomposition-based spatial rate control (DSRC) method.}
\centering
\scalebox{1}{
\begin{tabular}{c|c|c|c}
\toprule[1.5pt]
Model Index &RAATP & DSRC & BD-Rate (PSNR)\\ \hline
$M_1$&\XSolidBrush    &\XSolidBrush& 0.0\%  \\ \hline
$M_2$&\Checkmark    &\XSolidBrush  &--6.9\%\\ \hline
$M_3$&\Checkmark    &\Checkmark    & --10.7\%    \\
\bottomrule[1.5pt]
\end{tabular}
}
\label{effectiveness1}
\end{table}
%-------------------------------------------------------------------------
%-------------------------------------------------------------------------
\begin{figure}[t]
  \centering
   \includegraphics[width=\linewidth]{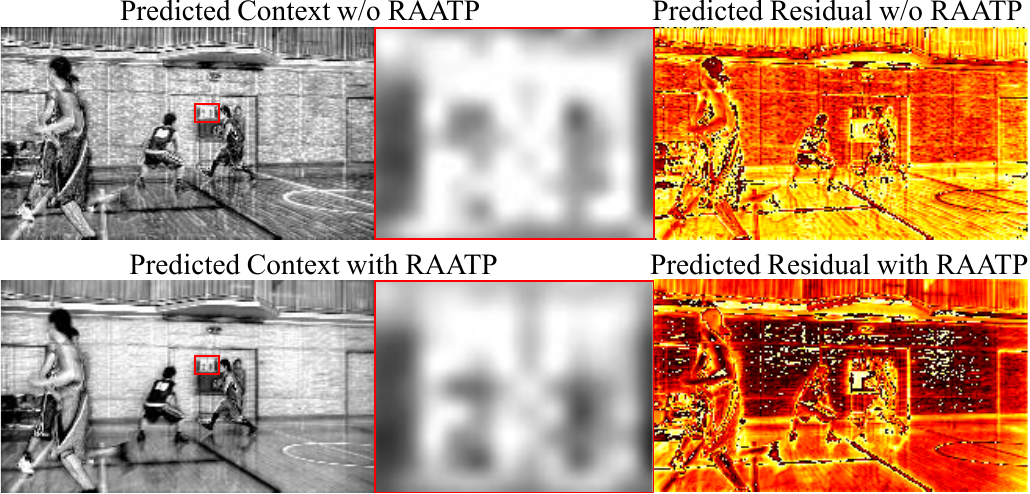}
      \caption{Visualization of the predicted temporal contexts and prediction residuals with and without our proposed rate-aware adaptive temporal prediction (RAATP) method. Darker colors indicate smaller residuals.}
   \label{fig:ablation_RAATP}
\end{figure}
%-------------------------------------------------------------------------

\subsection{Ablation Study}
\subsubsection{Effectiveness of Rate-Aware Adaptive Temporal Prediction}
To verify the effectiveness of the rate-aware adaptive temporal prediction (RAATP) method, we conduct an ablation study on the HEVC dataset. As shown in Table~\ref{effectiveness1}, starting from the baseline model \(M_1\) that employs neither RAATP nor the decomposition-based spatial rate control (DSRC) method, equipping RAATP alone in model \(M_2\) yields a BD‑rate saving of \(6.9\%\) in YUV420 PSNR. Fig.~\ref{fig:ablation_RAATP} offers a qualitative explanation for this gain by visualizing the predicted temporal contexts and the corresponding prediction residuals with and without RAATP. Without RAATP, the codec produces relatively smooth predictions that fail to capture fine motion details, leaving larger residuals in complex or occluded areas. When RAATP is enabled, the predicted context aligns more accurately with the current frame and the residuals are reduced.
%-------------------------------------------------------------------------
\begin{figure}[t]
  \centering
   \includegraphics[width=0.9\linewidth]{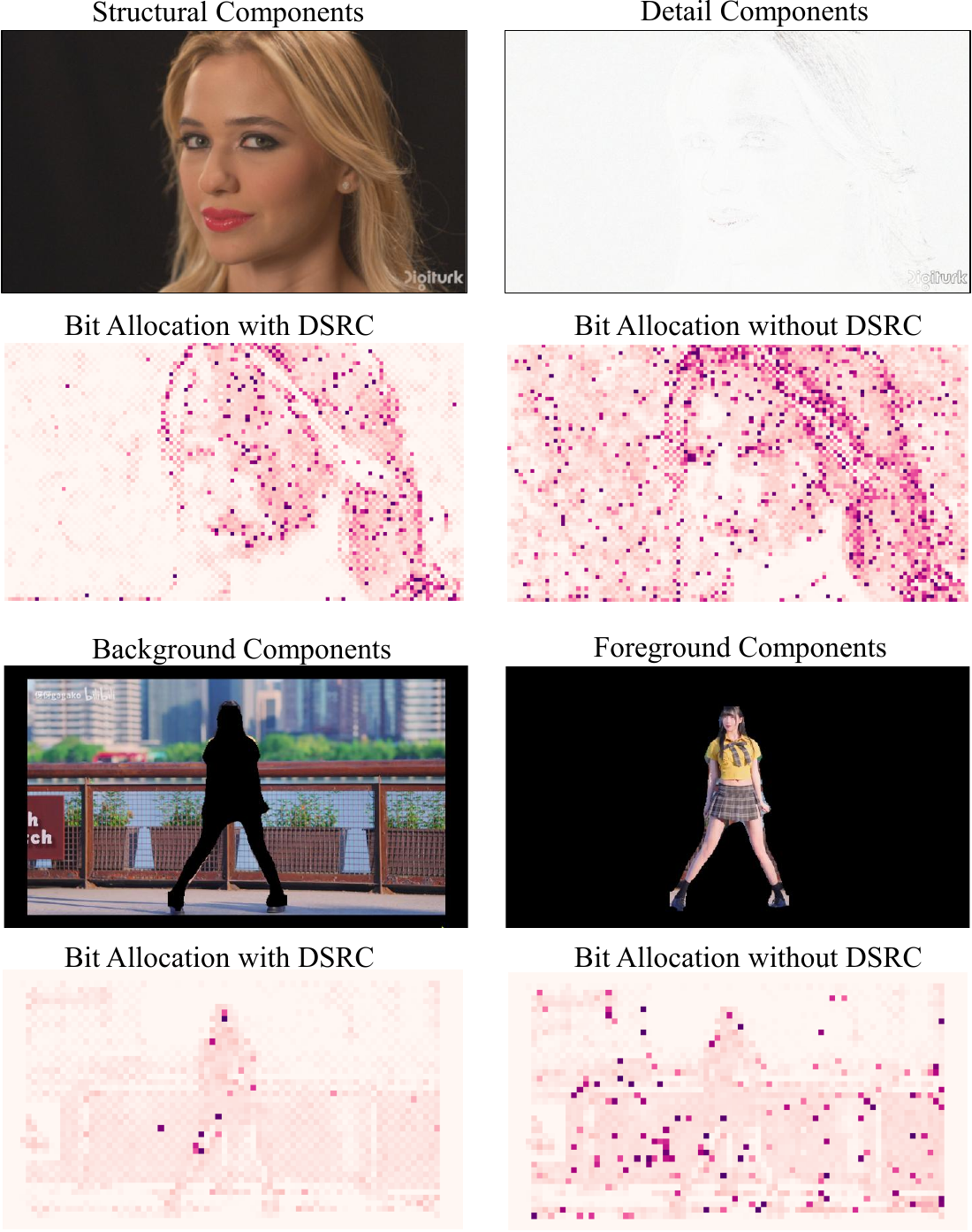}
      \caption{Illustration of the down‑up sampling-based decomposition and user-specified mask-based decomposition. Visualization of the bit allocation with and without our proposed decomposition-based spatial rate control (DSRC) method. Darker colors indicate more bits.}
   \label{fig:ablation_DSRC}
\end{figure}

%-------------------------------------------------------------------------
\subsubsection{Effectiveness of Decomposition-based Spatial Rate Control}
Table~\ref{effectiveness1} also quantifies the contribution of the  decomposition-based spatial rate control (DSRC) method. Adding DSRC on top of RAATP (from \(M_2\) to \(M_3\)) yields an additional BD‑rate saving of \(3.8\%\), confirming that spatial decomposition-based quantization further improves compression efficiency. Fig.~\ref{fig:ablation_DSRC} provides a visual interpretation of how DSRC achieves this gain. The first row shows the result of the down‑up sampling-based decomposition. The structural component appears smoother than the original frame, resembling a low‑pass filtered version that captures the main structures of the video, while the detail component retains the fine textures and edges. The second row compares the bit allocation maps with and without DSRC. Without DSRC, the codec distributes bits more uniformly across the textured and flat regions. When DSRC is enabled, the codec can adaptively allocate fewer bits to flat regions and more bits to texture-rich regions, leading to an efficient use of rate budgets.

A further benefit of this decomposition is that, at test time, the down‑up sampling operation in the encoder can be replaced by a user‑specified mask, enabling zero‑shot ROI coding. Since the detail quantization vector \(q_d^E\) has learned to preserve fine‑grained information and the structural quantization vector \(q_s^E\) has learned to encode smooth content at lower fidelity, substituting the mask naturally assigns higher reconstruction quality to the foreground regions while allowing the background to be more heavily quantized. As shown in Fig.~\ref{fig:ablation_DSRC}, taking the \textit{Dancer} sequence as an example, the user can specify the foreground to use the quantization vector corresponding to the high bitrate and the background to use that of the low bitrate. With DSRC, the user can effectively reduce the background bitrate without affecting the video generation quality.

 \begin{table}[t]
\caption{Effectiveness of proposed module bank-based perceptual switching (MBPS) method and frame generator-based perceptual switching method (FGPS).}
\centering
\scalebox{0.92}{
\begin{tabular}{c|c|c|c|c}
\toprule[1.5pt]
Model Index &MBPS & FGPS & BD-Rate (DISTS) & BD-Rate (LPIPS)\\ \hline
$M_4$&\XSolidBrush    &\XSolidBrush&0.0\%& 0.0\%  \\ \hline
$M_5$&\Checkmark    &\XSolidBrush  &--27.5\%&--16.9\% \\ \hline
$M_6$&\Checkmark    &\Checkmark    &--51.8\%& --44.1\%    \\
\bottomrule[1.5pt]
\end{tabular}
}
\label{effectiveness2}
\end{table}
%-------------------------------------------------------------------------

\subsubsection{Effectiveness of Perceptual Switching}
To verify the effectiveness of the proposed perceptual switching method, we conduct an ablation study that progressively enables its two key components: the module bank-based perceptual switching (MBPS), which replaces all quantization vectors and the factorized entropy model with their perceptual counterparts, and the frame generator-based perceptual switching (FGPS), which additionally replaces the frame generator with one trained for perceptual quality. We take our codec with signal‑fidelity mode as the baseline (\(M_4\)) and evaluate DISTS and LPIPS BD‑rate savings against it. As shown in Table~\ref{effectiveness2}, enabling MBPS alone in model \(M_5\) achieves BD‑rate savings of \(27.5\%\) in DISTS and \(16.9\%\) in LPIPS. This demonstrates that switching the module bank already yields a substantial perceptual improvement over the fidelity‑only baseline. When FGPS is further enabled in model \(M_6\), the BD‑rate savings increase to \(51.8\%\) in DISTS and \(44.1\%\) in LPIPS. The large additional gains indicate that a dedicated perceptual frame generator can further benefit the overall perceptual quality.

\section{Conclusion}\label{sec:conclusion}
In this paper, we propose a unified real-time neural video codec  with temporal, spatial, and perceptual adaptivity.
We first propose a rate‑aware adaptive temporal prediction method that generates diverse prediction candidates and couples candidate selection directly to bitrate constraints.
Second, we propose a decomposition‑based spatial rate control method that achieves finer‑grained spatial bit allocation and enables direct spatial rate control at test time.
Third, we propose a perceptual switching method that allows a codec to switch between a signal fidelity mode and a perceptual quality mode.
Extensive experiments validate that our codec delivers excellent objective quality, perceptual quality, and zero-shot ROI coding.
These results confirm that our methods significantly enhance the practicality of neural video coding, advancing it toward broader real-world deployment.

\bibliographystyle{ieeetr}
\bibliography{ref}

@inproceedings{blau2018perception,
  title={The perception-distortion tradeoff},
  author={Blau, Yochai and Michaeli, Tomer},
  booktitle={Proceedings of the IEEE conference on computer vision and pattern recognition},
  pages={6228--6237},
  year={2018}
}

@article{krogh1994neural,
  title={Neural network ensembles, cross validation, and active learning},
  author={Krogh, Anders and Vedelsby, Jesper},
  journal={Advances in neural information processing systems},
  volume={7},
  year={1994}
}

@inproceedings{li2023high,
  title={High visual-fidelity learned video compression},
  author={Li, Meng and Shi, Yibo and Wang, Jing and Huang, Yunqi},
  booktitle={Proceedings of the 31st ACM International Conference on Multimedia},
  pages={8057--8066},
  year={2023}
}

@article{ma2025diffusion,
  title={Diffusion-based perceptual neural video compression with temporal diffusion information reuse},
  author={Ma, Wenzhuo and Chen, Zhenzhong},
  journal={ACM Transactions on Multimedia Computing, Communications and Applications},
  volume={21},
  number={12},
  pages={1--22},
  year={2025},
  publisher={ACM New York, NY}
}

@article{ma2026diffvc,
  title={DiffVC-RT: Towards Practical Real-Time Diffusion-based Perceptual Neural Video Compression},
  author={Ma, Wenzhuo and Chen, Zhenzhong},
  journal={arXiv preprint arXiv:2601.20564},
  year={2026}
}

@inproceedings{mao2026generative,
  title={Generative neural video compression via video diffusion prior},
  author={Mao, Qi and Cheng, Hao and Yang, Tinghan and Jin, Libiao and Ma, Siwei},
  booktitle={Proceedings of the IEEE/CVF Conference on Computer Vision and Pattern Recognition},
  pages={43239--43248},
  year={2026}
}

@article{li2026progvc,
  title={ProGVC: Progressive-based Generative Video Compression via Auto-Regressive Context Modeling},
  author={Li, Daowen and Dong, Ruixiao and Chen, Ying and Li, Kai and Ding, Ding and Li, Li},
  journal={arXiv preprint arXiv:2603.17546},
  year={2026}
}

@article{qi2025generative,
  title={Generative latent coding for ultra-low bitrate image and video compression},
  author={Qi, Linfeng and Jia, Zhaoyang and Li, Jiahao and Li, Bin and Li, Houqiang and Lu, Yan},
  journal={IEEE Transactions on Circuits and Systems for Video Technology},
  year={2025},
  publisher={IEEE}
}

@article{sullivan2012overview,
  author={Sullivan, Gary J and Ohm, Jens-Rainer and Han, Woo-Jin and Wiegand, Thomas},
  title={{Overview of the high efficiency video coding (HEVC) standard}},
  journal={IEEE Transactions on Circuits and Systems for Video Technology},
  volume={22},
  number={12},
  pages={1649--1668},
  year={2012},
  publisher={IEEE}
}

@inproceedings{zhang2018unreasonable,
  title={The unreasonable effectiveness of deep features as a perceptual metric},
  author={Zhang, Richard and Isola, Phillip and Efros, Alexei A and Shechtman, Eli and Wang, Oliver},
  booktitle={Proceedings of the IEEE/CVF Conference on Computer Vision and Pattern Recognition (CVPR)},
  pages={586--595},
  year={2018}
}

@article{seedance2026seedance,
  title={Seedance 2.0: Advancing video generation for world complexity},
  author={Seedance, Team and Chen, De and Chen, Liyang and Chen, Xin and Chen, Ying and Chen, Zhuo and Chen, Zhuowei and Cheng, Feng and Cheng, Tianheng and Cheng, Yufeng and others},
  journal={arXiv preprint arXiv:2604.14148},
  year={2026}
}

@article{hu2022fvc,
  title={Fvc: An end-to-end framework towards deep video compression in feature space},
  author={Hu, Zhihao and Xu, Dong and Lu, Guo and Jiang, Wei and Wang, Wei and Liu, Shan},
  journal={IEEE Transactions on Pattern Analysis and Machine Intelligence},
  volume={45},
  number={4},
  pages={4569--4585},
  year={2022},
  publisher={IEEE}
}

@article{ding2020image,
  title={Image quality assessment: Unifying structure and texture similarity},
  author={Ding, Keyan and Ma, Kede and Wang, Shiqi and Simoncelli, Eero P},
  journal={IEEE Transactions on Pattern Analysis and Machine Intelligence},
  volume={44},
  number={5},
  pages={2567--2581},
  year={2020},
  publisher={IEEE}
}

@article{lu2020end,
  title={An end-to-end learning framework for video compression},
  author={Lu, Guo and Zhang, Xiaoyun and Ouyang, Wanli and Chen, Li and Gao, Zhiyong and Xu, Dong},
  journal={IEEE Transactions on Pattern Analysis and Machine Intelligence},
  year={2020},
  publisher={IEEE}
}

@inproceedings{mercat2020uvg,
  title={{UVG} dataset: 50/120fps 4K sequences for video codec analysis and development},
  author={Mercat, Alexandre and Viitanen, Marko and Vanne, Jarno},
  booktitle={Proceedings of the 11th ACM Multimedia Systems Conference},
  pages={297--302},
  year={2020}
}

@inproceedings{wang2016mcl,
  title={{MCL-JCV}: a {JND}-based {H.264/AVC} video quality assessment dataset},
  author={Wang, Haiqiang and Gan, Weihao and Hu, Sudeng and Lin, Joe Yuchieh and Jin, Lina and Song, Longguang and Wang, Ping and Katsavounidis, Ioannis and Aaron, Anne and Kuo, C-C Jay},
  booktitle={2016 IEEE International Conference on Image Processing (ICIP)},
  pages={1509--1513},
  year={2016},
  organization={IEEE}
}

@inproceedings{agustsson2020scale,
  title={Scale-space flow for end-to-end optimized video compression},
  author={Agustsson, Eirikur and Minnen, David and Johnston, Nick and Balle, Johannes and Hwang, Sung Jin and Toderici, George},
  booktitle={Proceedings of the IEEE/CVF Conference on Computer Vision and Pattern Recognition {(CVPR)}},
  pages={8503--8512},
  year={2020}
}

@inproceedings{rippel2019learned,
  title={Learned video compression},
  author={Rippel, Oren and Nair, Sanjay and Lew, Carissa and Branson, Steve and Anderson, Alexander G and Bourdev, Lubomir},
  booktitle={Proceedings of the IEEE/CVF International Conference on Computer Vision (ICCV)},
  pages={3454--3463},
  year={2019}
}

@inproceedings{cheng2019learning,
  title={Learning image and video compression through spatial-temporal energy compaction},
  author={Cheng, Zhengxue and Sun, Heming and Takeuchi, Masaru and Katto, Jiro},
  booktitle={Proceedings of the IEEE/CVF Conference on Computer Vision and Pattern Recognition (CVPR)},
  pages={10071--10080},
  year={2019}
}

@inproceedings{liu2020conditional,
  title={Conditional entropy coding for efficient video compression},
  author={Liu, Jerry and Wang, Shenlong and Ma, Wei-Chiu and Shah, Meet and Hu, Rui and Dhawan, Pranaab and Urtasun, Raquel},
  booktitle={European Conference on Computer Vision (ECCV)},
  pages={453--468},
  year={2020},
  organization={Springer}
}

@article{sheng2025prediction,
  title={Prediction and reference quality adaptation for learned video compression},
  author={Sheng, Xihua and Li, Li and Liu, Dong and Li, Houqiang},
  journal={IEEE Transactions on Image Processing},
  year={2025},
  publisher={IEEE}
}

@article{sheng2025bi,
  title={Bi-Directional Deep Contextual Video Compression},
  author={Sheng, Xihua and Li, Li and Liu, Dong and Wang, Shiqi},
  journal={IEEE Transactions on Multimedia},
  year={2025},
  publisher={IEEE}
}

@article{liu2022end,
  title={End-to-end neural video coding using a compound spatiotemporal representation},
  author={Liu, Haojie and Lu, Ming and Chen, Zhiqi and Cao, Xun and Ma, Zhan and Wang, Yao},
  journal={IEEE Transactions on Circuits and Systems for Video Technology},
  volume={32},
  number={8},
  pages={5650--5662},
  year={2022},
  publisher={IEEE}
}

@article{bross2021overview,
  title={Overview of the versatile video coding ({VVC}) standard and its applications},
  author={Bross, Benjamin and Wang, Ye-Kui and Ye, Yan and Liu, Shan and Chen, Jianle and Sullivan, Gary J and Ohm, Jens-Rainer},
  journal={IEEE Transactions on Circuits and Systems for Video Technology},
  year={2021},
  publisher={IEEE}
}

@article{xue2019video,
  title={Video enhancement with task-oriented flow},
  author={Xue, Tianfan and Chen, Baian and Wu, Jiajun and Wei, Donglai and Freeman, William T},
  journal={International Journal of Computer Vision},
  volume={127},
  number={8},
  pages={1106--1125},
  year={2019},
  publisher={Springer}
}

@article{kingma2014adam,
  title={Adam: A method for stochastic optimization},
  author={Kingma, Diederik P and Ba, Jimmy},
  journal={arXiv preprint arXiv:1412.6980},
  year={2014}
}

@article{yilmaz2021end,
  title={End-to-End Rate-Distortion Optimized Learned Hierarchical Bi-Directional Video Compression},
  author={Y{\i}lmaz, M Ak{\i}n and Tekalp, A Murat},
  journal={IEEE Transactions on Image Processing},
  volume={31},
  pages={974--983},
  year={2021},
  publisher={IEEE}
}

@article{sheng2022temporal,
  title={Temporal context mining for learned video compression},
  author={Sheng, Xihua and Li, Jiahao and Li, Bin and Li, Li and Liu, Dong and Lu, Yan},
  journal={IEEE Transactions on Multimedia},
  volume={25},
  pages={7311--7322},
  year={2022},
  publisher={IEEE}
}

@inproceedings{ho2022canf,
  title={Canf-vc: Conditional augmented normalizing flows for video compression},
  author={Ho, Yung-Han and Chang, Chih-Peng and Chen, Peng-Yu and Gnutti, Alessandro and Peng, Wen-Hsiao},
  booktitle={Computer Vision--ECCV 2022: 17th European Conference, Tel Aviv, Israel, October 23--27, 2022, Proceedings, Part XVI},
  pages={207--223},
  year={2022},
  organization={Springer}
}

@article{chen2024maskcrt,
  title={Maskcrt: Masked conditional residual transformer for learned video compression},
  author={Chen, Yi-Hsin and Xie, Hong-Sheng and Chen, Cheng-Wei and Gao, Zong-Lin and Benjak, Martin and Peng, Wen-Hsiao and Ostermann, J{\"o}rn},
  journal={IEEE Transactions on Circuits and Systems for Video Technology},
  year={2024},
  publisher={IEEE}
}

@article{lin2022dmvc,
  title={{DMVC}: Decomposed motion modeling for learned video compression},
  author={Lin, Kai and Jia, Chuanmin and Zhang, Xinfeng and Wang, Shanshe and Ma, Siwei and Gao, Wen},
  journal={IEEE Transactions on Circuits and Systems for Video Technology},
  year={2022},
  publisher={IEEE}
}

@article{guo2023learning,
  title={Learning Cross-Scale Weighted Prediction for Efficient Neural Video Compression},
  author={Guo, Zongyu and Feng, Runsen and Zhang, Zhizheng and Jin, Xin and Chen, Zhibo},
  journal={IEEE Transactions on Image Processing},
  year={2023},
  publisher={IEEE}
}

@inproceedings{li2023neural,
  title={Neural video compression with diverse contexts},
  author={Li, Jiahao and Li, Bin and Lu, Yan},
  booktitle={Proceedings of the IEEE/CVF Conference on Computer Vision and Pattern Recognition (CVPR)},
  pages={22616--22626},
  year={2023}
}

@article{sheng2024vnvc,
  title={VNVC: A Versatile Neural Video Coding Framework for Efficient Human-Machine Vision},
  author={Sheng, Xihua and Li, Li and Liu, Dong and Li, Houqiang},
  journal={IEEE Transactions on Pattern Analysis and Machine Intelligence},
  year={2024},
  publisher={IEEE}
}

@article{sheng2024spatial,
  title={Spatial Decomposition and Temporal Fusion based Inter Prediction for Learned Video Compression},
  author={Sheng, Xihua and Li, Li and Liu, Dong and Li, Houqiang},
  journal={IEEE Transactions on Circuits and Systems for Video Technology},
  year={2024},
  publisher={IEEE}
}

@inproceedings{li2024neural,
  title={Neural video compression with feature modulation},
  author={Li, Jiahao and Li, Bin and Lu, Yan},
  booktitle={Proceedings of the IEEE/CVF Conference on Computer Vision and Pattern Recognition},
  pages={26099--26108},
  year={2024}
}

@inproceedings{radford2021learning,
  title={Learning transferable visual models from natural language supervision},
  author={Radford, Alec and Kim, Jong Wook and Hallacy, Chris and Ramesh, Aditya and Goh, Gabriel and Agarwal, Sandhini and Sastry, Girish and Askell, Amanda and Mishkin, Pamela and Clark, Jack and others},
  booktitle={International conference on machine learning},
  pages={8748--8763},
  year={2021},
  organization={PMLR}
}

@article{wei2025rdvc,
  title={RDVC: Efficient Deep Video Compression with Regulable Rate and Complexity Optimization},
  author={Wei, Xiaojie and Lin, Jielian and Xu, Jiawei and Gao, Wei and Zhao, Tiesong},
  journal={IEEE Transactions on Multimedia},
  year={2025},
  publisher={IEEE}
}

@article{yuan2025mining,
  title={Mining Temporal Redundancy Using Long Short-Term Motion Aggregation and Global-Local Decorrelation for Learned Video Compression},
  author={Yuan, Feng and Pan, Zhaoqing and Lei, Jianjun and Peng, Bo and Xie, Haoran and Wang, Fu Lee and Kwong, Sam},
  journal={IEEE Transactions on Circuits and Systems for Video Technology},
  year={2025},
  publisher={IEEE}
}

@inproceedings{jia2025towards,
  title={Towards practical real-time neural video compression},
  author={Jia, Zhaoyang and Li, Bin and Li, Jiahao and Xie, Wenxuan and Qi, Linfeng and Li, Houqiang and Lu, Yan},
  booktitle={Proceedings of the Computer Vision and Pattern Recognition Conference},
  pages={12543--12552},
  year={2025}
}

@inproceedings{jiang2025ecvc,
  title={Ecvc: Exploiting non-local correlations in multiple frames for contextual video compression},
  author={Jiang, Wei and Li, Junru and Zhang, Kai and Zhang, Li},
  booktitle={Proceedings of the Computer Vision and Pattern Recognition Conference},
  pages={7331--7341},
  year={2025}
}

@inproceedings{yang2022perceptual,
  title={Perceptual Learned Video Compression with Recurrent Conditional GAN.},
  author={Yang, Ren and Timofte, Radu and Van Gool, Luc},
  booktitle={IJCAI},
  pages={1537--1544},
  year={2022}
}

@inproceedings{fathima2023neural,
  title={A neural video codec with spatial rate-distortion control},
  author={Fathima, Noor and Petersen, Jens and Sauti{\`e}re, Guillaume and Wiggers, Auke and Pourreza, Reza},
  booktitle={Proceedings of the IEEE/CVF Winter Conference on Applications of Computer Vision},
  pages={5365--5374},
  year={2023}
}

@inproceedings{wu2024roi,
  title={Roi-dvc: A region-of-interest based deep video coding framework},
  author={Wu, Xiaojie and Wang, Ping and Wang, Xinhong},
  booktitle={2024 IEEE International Conference on Image Processing (ICIP)},
  pages={1967--1972},
  year={2024},
  organization={IEEE}
}

@inproceedings{liu2024roi,
  title={ROI-aware dynamic network quantization for neural video compression},
  author={Liu, Jiamin and Zhang, Baochang and Cao, Xianbin},
  booktitle={International Conference on Pattern Recognition},
  pages={333--349},
  year={2024},
  organization={Springer}
}

@article{chen2003roi,
  title={ROI video coding based on H. 263+ with robust skin-color detection technique},
  author={Chen, Mei-Juan and Chi, Ming-Chieh and Hsu, Ching-Ting and Chen, Jeng-Wei},
  journal={IEEE Transactions on Consumer Electronics},
  volume={49},
  number={3},
  pages={724--730},
  year={2003},
  publisher={IEEE}
}
\end{document}